\documentclass[a4paper,11pt]{article}
\usepackage{jheppub}
\notoc

\title{A Very Intense Neutrino Super Beam Experiment for Leptonic CP Violation Discovery based on the European Spallation Source Linac: A Snowmass 2013 White Paper}

\author[m]{E.~ Baussan,}
\author[l]{M.~ Blennow,}
\author[k]{M.~Bogomilov,} 
\author[m]{E.~ Bouquerel,}
\author[f]{J.~ Cederk\"all,}
\author[f]{P.~ Christiansen,}
\author[b]{P.~ Coloma,}
\author[e]{P.~ Cupial,}
\author[g]{H.~Danared,} 
\author[c]{C.~Densham,} 
\author[m,*]{M.~Dracos,}
\author[n,*]{T.~Ekel\"of,}
\author[g]{M.~Eshraqi,}
\author[h]{E.~Fernandez~Martinez,}
\author[m]{G.~ Gaudiot,}
\author[g]{R.~Hall--Wilton,}
\author[n,d]{J.--P.~Koutchouk,} 
\author[g]{M.~Lindroos,} 
\author[k]{R.~Matev,} 
\author[g]{D.~McGinnis,} 
\author[j]{M.~Mezzetto,} 
\author[g]{R.~Miyamoto,} 
\author[i]{L.~Mosca,} 
\author[l]{T.~Ohlsson,} 
\author[n]{H.~\"Ohman,} 
\author[m]{F.~ Osswald,}
\author[g]{S.~Peggs,} 
\author[m]{P.~Poussot,} 
\author[n]{R.~Ruber,}
\author[a]{J.Y.~Tang,} 
\author[k]{R.~Tsenov,} 
\author[k]{G.~Vankova--Kirilova,} 
\author[m]{N.~Vassilopoulos,}
\author[d]{E.~Wildner,}
\author[m]{and J.~Wurtz.}

\affiliation[a]{Institute of High Energy Physics, CAS, Beijing 100049, China.}
\affiliation[b]{Center for Neutrino Physics, Virginia Tech, Blacksburg, VA 24061, USA.}
\affiliation[c]{STFC Rutherford Appleton Laboratory, OX11 0QX Didcot, UK.}
\affiliation[d]{CERN, CH-1211, Geneva 23, Switzerland.}
\affiliation[e]{AGH University of Science and Technology, Al. Mickiewicza 30, 30-059 Krakow, Poland.}
\affiliation[f]{Department of Physics, Lund University, Box 118, SE--221 00 Lund, Sweden.}
\affiliation[g]{European Spallation Source, ESS AB, P.O Box 176, SE-221 00 Lund, Sweden.}
\affiliation[h]{Dpto. de F\'isica T\'eorica and Instituto de F\'isica T\'eorica UAM/CSIC, Universidad Aut\'onoma de Madrid, Cantoblanco E-28049 Madrid, Spain.}
\affiliation[i]{Laboratoire Souterrain de Modane, F-73500 Modane, France.}
\affiliation[j]{INFN Sezione di Padova, 35131 Padova, Italy.}
\affiliation[k]{Department of Atomic physics, St. Kliment Ohridski University of Sofia, Sofia, Bulgaria.}
\affiliation[l]{Department of Theoretical Physics, School of Engineering Sciences, KTH Royal Institute of Technology, AlbaNova University Center, SE-106 91 Stockholm, Sweden.}
\affiliation[m]{IPHC, Universit\'e de Strasbourg, CNRS/IN2P3, F-67037 Strasbourg, France}
\affiliation[n]{Department of  Physics and Astronomy, Uppsala University, Box 516, SE-75120 Uppsala, Sweden}

\affiliation[*]{Corresponding Authors: marcos.dracos@in2p3.fr and tord.ekelof@physics.uu.se}

\abstract{

Very intense neutrino beams and large neutrino detectors will be needed in order to enable the discovery of CP violation in the leptonic sector.
We propose to use the proton linac of the European Spallation Source currently under construction in Lund, Sweden to deliver, in parallel with the spallation neutron production, a very intense, cost effective and high performance neutrino beam.
The baseline program for the European Spallation Source linac is that it will be fully operational at 5~MW average power by 2022, producing 2~GeV 2.86~ms long proton pulses at a rate of 14~Hz.
Our proposal is to upgrade the linac to 10~MW average power and 28~Hz, producing 14 pulses/s for neutron production and 14~pulses/s for neutrino production.
Furthermore, because of the high current required in the pulsed neutrino horn, the length of the pulses used for neutrino production needs to be compressed to a few $\mu$s with the aid of an accumulator ring.
A long baseline experiment using this Super Beam and a megaton underground Water Cherenkov detector located in existing mines 300--600~km from Lund will make it possible to discover leptonic CP violation at 5~$\sigma$ significance level in  up to 50\% of the leptonic Dirac CP--violating phase range.
This experiment could also determine the neutrino mass hierarchy at a significance level of more than 3~$\sigma$ if this issue will not already have been settled by other experiments by then.
The mass hierarchy performance could be increased by combining the neutrino beam results with those obtained from atmospheric neutrinos detected by the same large volume detector.
This detector will also be used to measure the proton lifetime, detect cosmological neutrinos and neutrinos from supernova explosions. Results on the sensitivity to leptonic CP violation and the neutrino mass hierarchy are presented.
}

\begin{document}
\maketitle

\section{Introduction}

2012 was an important year for neutrino physics as the last unknown leptonic mixing angle $\theta_{13}$ was measured and found to be, not only different from zero, but also saturating previous bounds~\cite{1106.2822, 1112.6353, 1204.0626, 1203.1669}.
This opens the door to the experimental observation of possible CP violation in the leptonic sector using classical neutrino beams, an observation that will have important cosmological implications.
CP violation is a necessary ingredient to generate the observed matter dominance in the universe and it has been shown that the measured  amount of CP violation in the quark sector is not enough to account for it~\cite{hep-ph/9312215, hep-ph/9406289}.
Previous feasibility studies of experiments, which were aimed at being able to measure as low as possible values of $\theta_{13}$ now have to readjust their parameters to optimize their performance  for CP violation and mass hierarchy measurements given the measured large $\theta_{13}$ value.
Moreover, a value of large $\theta_{13}$ enhances the physics performance of experiments with the detector placed at the second oscillation maximum as compared to those with the detector placed at the first oscillation maximum~\cite{1110.4583}.

The ESS$\nu$SB (standing for European Spallation Source Neutrino Super Beam) project succeeds the studies made by the FP7 Design Study EURO$\nu$~\cite{euronu-url, 1305.4067, 1212.0732},  regarding future neutrino facilities, in particular the study made of the CERN based Superconducting Proton Linac (SPL)~\cite{CERN-2006-06, Brunner:2008zza}  (4.5~GeV protons, 4~MW) Super Beam and the MEMPHYS~\cite{hep-ex/0607026, 1206.6665}  large Water Cherenkov detector in the Fr\'ejus tunnel located at the first neutrino oscillation maximum (130 km).
ESS$\nu$SB~\cite{1212.5048} proposes to study a Super Beam, which uses the high power linac of the European Spallation Source (ESS)~\cite{ess-url}  at Lund in Sweden as proton driver and with a MEMPHYS type detector located in a deep mine at a distance between 300~km to 600~km distance, near the second neutrino oscillation maximum.
This proposal is similar to the SPL based Super Beam proposal which is used as reference.

The ESS$\nu$SB  presents many synergies with the ESS primary purpose, which is the production and use of spallation neutrons.
The proposal is to use the ESS proton driver (2.0 GeV protons, 5 MW) simultaneously for the two applications with no reduction in the spallation neutron production, thus decreasing considerably the cost of the proposed project as compared to constructing a dedicated proton driver.
In a second stage, the Neutrino Factory (NF)~\cite{Apollonio:2008aa}, considered as the ultimate neutrino facility, could make use of this proton driver.
The 2.86~ms long pulses of the ESS linac would need to be reduced to a few $\mu$s long pulses allowing a limitation of the length of the very high current pulse in the hadronic collector (horn) producing severe heat dissipation problems. 
In order to do so, H$^-$ ions have to be accelerated in the linac and injected in a proton accumulator ring to be designed for this application.
Detailed studies will be made of the modifications of the ESS proton linac required to allow simultaneous acceleration of H$^+$ and H$^-$ ions at an average power of 5+5~MW.
In particular, it is important to identify modifications that can be made without high cost already in the build--up stage of the linac so as to facilitate and reduce the cost of the overall linac modification in case this modification would have to be made only after the completion of the linac.

A considerable fraction of the future spallation neutron users at ESS are also in need of short pulses and the costs of the H$^-$ beam and the accumulator ring could therefore be shared.
This highly synergetic beam and ring will be extensively studied in this project with the aim to design a common facility satisfying the requirements for both the short pulsed neutron measurements and the neutrino measurements.

The EURO$\nu$ studies identified some key elements of the SPL Super Beam for which further R\&D would be necessary like the proton target, the hadron collector and the collector pulse generator.
The corresponding items for the ESS based neutrino project will be studied to prove their feasibility.

An important part of the study is to evaluate through simulations the optimal distance of the detector from the neutrino source.
Once the result of this optimization has been made, the mine in Sweden that is at a distance closest to the optimum will be investigated in detail with regard to where in the mine to excavate the MEMPHYS underground halls and what shape and methods of construction should be chosen for the halls.
ESS$\nu$SB will profit from the studies already performed in the framework of the FP7 LAGUNA Design Study~\cite{laguna-url}.

\section{The ESS Facility}

ESS will be a major user facility providing slow neutrons for research laboratories and industry.
A first proton beam for neutron production will be delivered at reduced energy and power by 2019.
A proton beam of the full design power 5~MW and energy 2.0~GeV will be delivered by 2022.
There will be 14 pulses of 62.5~mA current and 2.86~ms length per second.
ESS is supported by the European Strategy Forum for Research Infrastructures (ESFRI).

Protons will be sent to a rotating tungsten target. 
The generated neutrons will be detected by 22 neutron scattering research instruments.
The annual operation period will be 5000~hours (208 days, $1.8\times 10^7$ sec).
In table~\ref{tab1} the main ESS parameters are given.

\begin{table}[htdp]
\caption{Main ESS facility parameters concerning the proton beam.}
\begin{center}
\begin{tabular}{lc}
\hline
Parameter & Value \\
\hline
Average beam power & 5~MW \\
Proton kinetic energy & 2.0 GeV \\
Average macro--pulse current & 62.5~mA \\
Macro--pulse length & 2.86~ms \\
Pulse repetition rate & 14~Hz \\
Maximum accelerating cavity surface field & 45~MV/m \\
Maximum linac length (excluding contingency and upgrade space) & 352.5~m \\
Annual operating period & 5000~h \\
Reliability & 95\% \\
\hline
\end{tabular}
\end{center}
\label{tab1}
\end{table}%

\begin{figure}[hbt]
\begin{center}
\setlength\fboxsep{0.6pt}
\setlength\fboxrule{0.6pt}
  \includegraphics[width=1\textwidth]{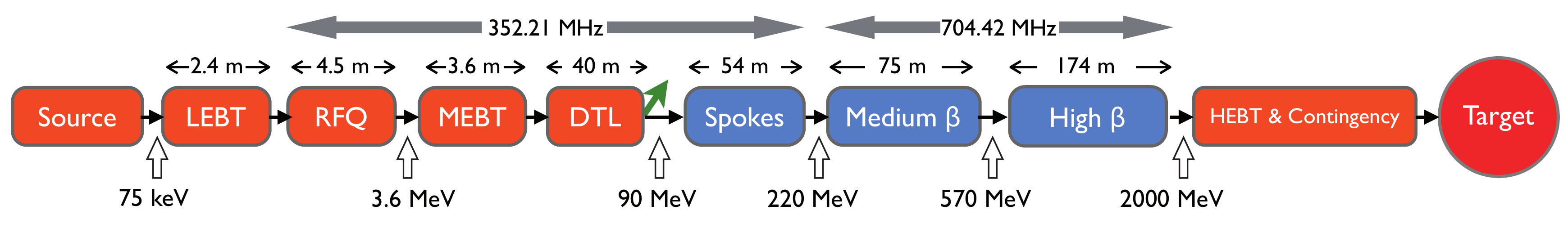}
\end{center}
\caption{  Schematic drawing of the 2.0~GeV proton linear accelerator of the European Spallation Source. Empty space for future upgrades is left at the end of the linac.}
\label{linac} 
\end{figure}

In order for the ESS to be used to generate a neutrino beam in parallel with the spallation neutrons, some modifications of the proton linac are necessary.
The current generating the magnetic field in the magnetic horn that focuses the produced hadrons downstream of the target needs to be so high that  maintaining it for more than a few microseconds would lead to excessive heating of the current conductor.
To focus all of the produced hadrons, the proton beam pulse must not be longer than the flat part of the current pulse in the horn.
The proton pulse length in the linac of 2.86~ms can be reduced to 1.5 ~$\mu$s by multiturn  injecting of the 2.86~ms proton pulse into a 426~m circumference accumulator ring  and then eject the stored protons in one turn.
Due to the difficulty to inject protons in the accumulator while a large amount of protons is already circulating in it, H$^-$ ions, instead of protons (H$^+$), need to be accelerated in the linac.
The H$^-$ ions will be stripped of their two electrons using a laser beam at the position where the linac beam enters the accumulator ring.
The currently planned ESS linac  (Fig.~\ref{linac}) gives an excellent opportunity to provide, simultaneously with proton acceleration for spallation neutron production, H$^-$ ion acceleration for the production of a uniquely high intensity neutrino beam, thus allowing for a new generation of neutrino beam experiments.
An important boundary condition is that the operation of the linac for neutrino beam production needs to be such that it does not in any respect downgrade the ESS capacity for spallation neutron production.

\section{The ESS Linac Modifications}

As already mentioned, the current design of the ESS linac will make possible the acceleration of 2.86~ms long 62.5~mA proton pulses to 2.0~GeV at 14~Hz. The proposed plan for simultaneous H$^-$ acceleration is to have one 2.86~Êms long 62.5~mA H$^-$ pulse accelerated in the 71.4~ms long gap between the proton (H$^+$) pulses, requiring the linac pulse frequency of 14~Hz to be raised to 28~Hz.
The radiofrequency (RF) is 352.2~MHz in the first low energy spoke cavity section of the ESS linac and 704.4~MHz in the second high--energy elliptical cavity section.
This implies that the RF phase of the second section would have to be dynamically shifted back and forth between the H$^+$ and H$^-$ pulses.

An H$^-$ source will have to be added to the H$^+$ source.
The optimum position along the linac at which the beams from the two sources are combined will be studied.
A duplication of the very low energy part of the linac will be needed probably up to some point downstream of the Radio Frequency Quadrupole (RFQ). 

The whole acceleration process of the H$^-$ beam will be studied using simulations.
Mechanisms possibly responsible for the emittance blow--up of the beams will be evaluated.
Blow--up is more critical for the H$^-$ beam, since it has to be injected in an accumulator ring, than for the H$^+$ beam, which will only have to be directed towards a spallation target.

Another phenomenon that will be studied is the stripping of the H$^-$ ions.
Possible Lorentz stripping of the H$^-$ ions which may occur in the fields of the magnets has to be studied, in particular in the H$^-$ beam switching to the accumulator and the injection into the accumulator before the laser stripping.
H$^-$ ions can also lose their electrons in collisions with residual gas and with blackbody photons, and by the so--called intrabeam stripping first observed at LEAR~\cite{Chanel:1987et} and identified as the primary source of beam losses at SNS (Spallation Neutron Source, USA)~\cite{sns-url, doi.108.114801}.

To avoid beam spillage at the injection into the ring  and during RF capture, one needs to introduce a gap in the train of bunches using a high--speed chopper.
Because of this chopping, the H$^-$ beam pulse will have spectral components within a few 10~MHz of the nominal frequencies and the generated harmonics may excite Higher Order Modes (HOMs) in the accelerating cavities.
These HOMs will be simulated in detail to determine an acceptable chopping scheme.
A careful collimation and shielding design, based on simulations, will be made with the aim to minimize beam losses.

The average power of the linac will have to be doubled from 5~MW to 10~MW, requiring a corresponding doubling of the average output power from the RF sources.
The cost for this doubling will be the dominant cost of the linac modification project.
If the power upgrade can be carried out already during the initial build--up of the linac, modulators, amplifiers and power transfer equipment need to be designed for the doubled average power of the linac.
If the power upgrade will have to wait to a later stage, the original RF power sources can either be made such that they can, at this later stage, be upgraded to twice the average power or the power sources would have to be doubled in number.
For the latter option it is important that there be enough free space available at the side of the linac to house the new power sources and technical services.

The effect of doubling the average linac power on the cooling of the cavities, of the power couplers and of the cryo--modules, will be studied.
A prototype 352~MHz spoke cavity for the ESS linac will be tested at Uppsala University in Sweden already as from July 2014 in a cryostat at 14~Hz pulse frequency and at the full instantaneous power required for ESS proton acceleration, which is 350~kW.
As source for the 352~MHz power both a tetrode amplifier and a solid state amplifier will be tested.
As part of the ESS$\nu$SB project, the power supplied will be doubled and the pulse frequency raised to 28~Hz, thus doubling the average power to the cavity.
The influence of this higher power on the operation of the cavity and on the capacity to cool the cavity itself and, in particular, its RF coupler will be studied.
As to the 704~MHz modulator and klystron power source planned for the medium beta section of the linac it will be possible to order a prototype of a so--called variable charging--rate modulator which will allow tests at 28~Hz pulse frequency.
As to the technology planned for the RF source of the high beta section of the linac, the current plan is to investigate the possibility of using Inductive Output Tubes. The way to increase their pulse frequency to 28 Hz will also be studied.

\section{The Accumulator Ring}

As already explained it will be important to have as short as possible neutrino pulses in order to minimize the duration of the current pulses sent to the hadron collector.
In this way, the background from cosmic rays in the large MEMPHYS detector, already low due to the depth ($\sim$1000~m) of the detector location, is reduced too.

A first study of a 318~m circumference accumulator ring to compress the pulses has already been made in the EURO$\nu$ Super Beam project.
This study  will be continued in ESS$\nu$SB.
Each pulse from the ESS linac will contain $1.1\times 10^{15}$ protons, which for a normalized beam emittance (95\%) of 100-mm-mrad in the ring by multi-turn injection
(the emittance from the linac should be in the order of a few mm-mrad) will lead to the very large space--charge tune--shift of about 0.75.
A way to reduce the tune shift is to divide the ring up on 4 superposed rings located in the same tunnel, each ring receiving 1/4 of the bunches during the multi--turn injection.
This will lead to a reduction of the tune shift to the level of around 0.2, which is acceptable for the 2.86~ms storage time.

There has to be enough space between the bunches in the bunch train from the linac to permit the beam distribution system to inject from one ring to the next one.
Experience already exists from the CERN PS Booster \cite{psb-url} of using 4 superimposed rings with the aim to avoid high space charge effects.
The lattices and collimators of a single ring accumulator and of a four rings accumulator will be designed.
These designs will be used for simulations of the accumulation process in which high order momentum compaction factor will be taken into account, in order to evaluate collective effects and to develop optimization and correction schemes. 

As described in Section~\ref{target}, four separate targets are needed in order to mitigate the high power dissipation in the target material.
Each of the four beams from the four accumulator rings will be led to one of the targets.
Within EURO$\nu$, a beam distribution system downstream of a single accumulator ring to four target stations has already been studied~\cite{1212.0732}.
For ESS$\nu$SB a similar system will be studied for the distribution of the beam from the linac to the four rings.

The H$^-$ ions will be fully stripped at injection into the accumulator using a laser--stripping device (foil stripping could not be used as the foils would not resist the high beam power).
As already mentioned, an empty space in the bunch train will be provided in the linac by chopping the beam regularly according to the circumference of the accumulator (multi--turn injection).
The extraction of the beam from the ring needs a group of kickers which should have a rise time of not more than 100~ns.

The injection switchyard, the multi--turn injection process, the stripping system, the extraction system (including kickers and septum magnet) and the system to guide the beams from the four accelerator rings to the four targets will be developed using simulations.

The possibility to make laser--stripping experimental tests using existing set--ups in the USA~\cite{Nesterenko:2007zza, doi.108.114801} will be investigated.
The simulation studies need to include shielding and collimation studies which should be connected to those done for the linac itself.
The beam instrumentation needed to monitor and steer the beams in the accumulator rings will also be studied.

The physics performance simulations already show that a somewhat higher proton energy than 2.0~GeV, like 2.5~GeV or 3.0~GeV, would be advantageous from the physics point of view.
Therefore, the possibilities to use the accumulator to also accelerate the beam to a higher energy will be investigated.
It may be noted that there is about 100~m empty space in the end of the linac tunnel intended  for future energy upgrades allowing to go up to 3.0~GeV proton kinetic energy by prolonging the linac.

\begin{figure}[hbt]
\begin{center}
\includegraphics[width=0.7\textwidth]{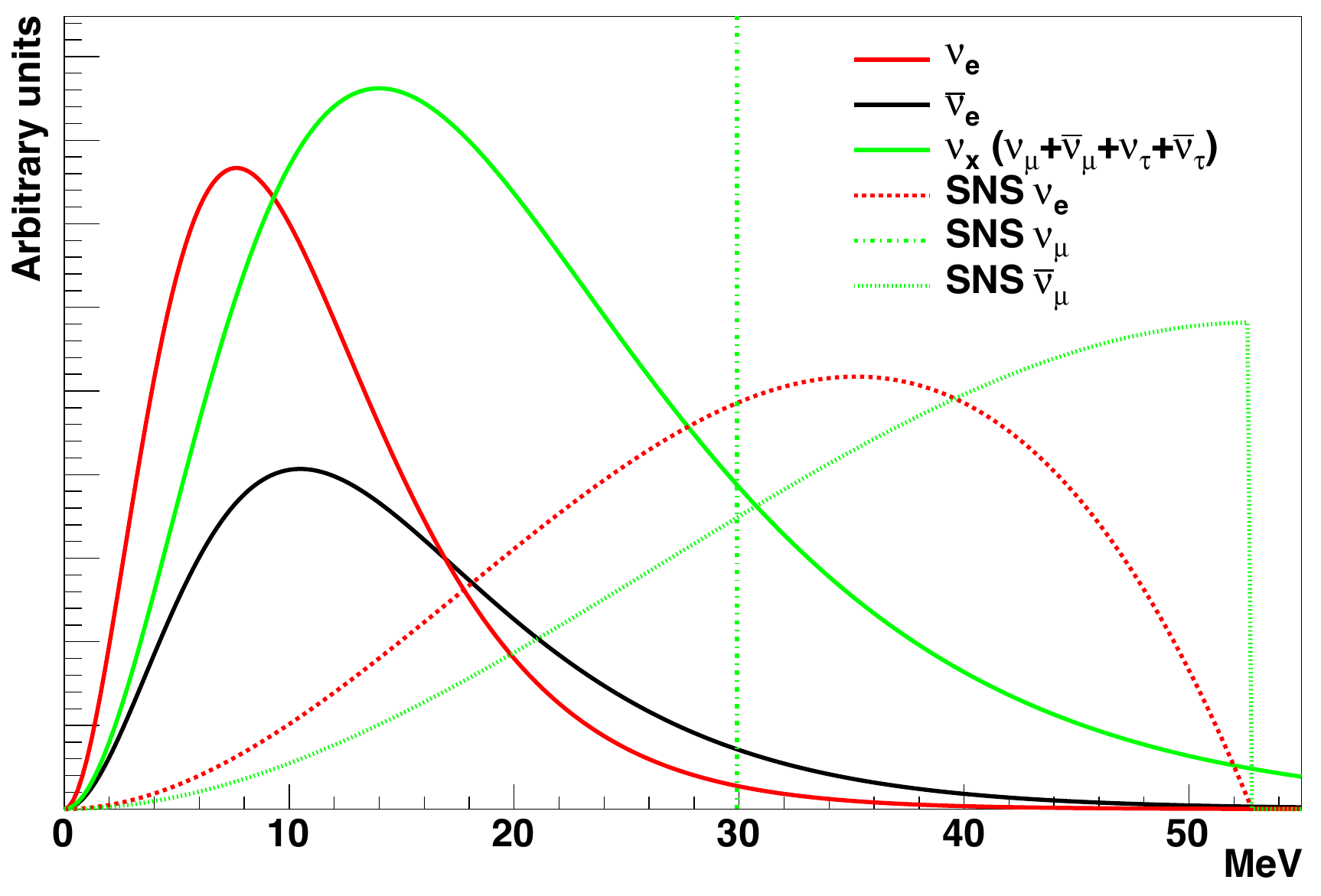}
\end{center}
\caption{Typical expected supernova neutrino spectrum for different flavors (solid lines) and SNS neutrino spectrum (dashed and dotted lines).}
\label{snsfig}
\end{figure}

The reduction of the proton pulse duration would also open the door to other types of neutrino experiments based on the neutron spallation facility.
Several such experiments have been proposed for the SNS facility.
One experiment is to use neutrinos produced by pions and muons decaying at rest in the neutron spallation target for neutrino oscillation studies.
Another is to measure neutrino cross--sections, in particular those interesting for supernova neutrino measurements.
The $\pi^+$ decay at rest and subsequent $\mu^+$ decay also at rest leading to the production of $\nu_\mu$, $\nu_e$ and $\bar\nu_\mu$.
The neutrinos produced in pion decaying at rest are mono--energetic (Fig.~\ref{snsfig}).
These neutrinos can be used for high--precision neutrino experiments. 
Multi-purpose experiments are already proposed~\cite{1211.5199} that will perform a search for light sterile neutrinos, searches for beyond the Standard Model interactions based on neutrino oscillations, and provide tests of Standard Model predictions through precision neutrino cross section measurements.
These experiments require the proton pulses to be no longer than a few $\mu$s because of the high background coming from the facility itself (mainly from radioactivity) or from cosmic rays.
The provision of short proton pulses also makes possible the separation of neutrinos coming from pion decays and those coming from the subsequent muon decays.
At ESS, having about five times proton beam power than SNS, these experiments would have higher event statistics.
Specific proposals for ESS have been presented at the workshop ``Neutrino, Neutron, Nuclear, Medical and Muon Physics at ESS" held in 2009~\cite{3n2mp-url}.

It is of significant interest to note that there are spallation neutron users of the ESS who have expressed a keen interest in having shorter pulses than 2.86~ms.
For neutrons, the optimal pulse length would be of the order of 100~$\mu s$.
This will significantly enhance the neutron peak flux compared to the present design~\cite{mats-ipac}.
It would not be possible, for economical reasons, to have an accumulator ring with a circumference large enough (of order 30~km) to produce such long pulses by single turn extraction.
However, using multiturn extraction much smaller accumulator rings can be used, such as those  proposed for the ESS$\nu$SB project.
This option has already been discussed within the ESS and the synergy between the two uses of the accumulator opens the perspective of sharing the investment and operation costs for the H$^-$ beam and the accumulator with the spallation neutron users.

Finally it may be noted that the powerful proton driver and accumulator based on the ESS linac as proposed above could possibly, at a later stage, be used for the realization of a Neutrino Factory~\cite{Apollonio:2008aa}, which would allow to substantially enhance the performance for neutrino measurements.

\section{The Target Station}
\label{target}

The target station includes the target itself that is hit by the protons leading to the production of short lived mesons, mainly pions, which decay producing muon neutrinos. 
Other main components of the target station are the hadron collector called magnetic horn, which focuses the hadrons towards the far neutrino detector, and the decay tunnel, long enough to allow the mesons to decay, but not as long as to allow for a significant amount of the muons to decay.

In order to mitigate the detrimental effects of the very high power of the proton beam hitting the target, EURO$\nu$~\cite{euronu-url, 1305.4067} has proposed a system with four targets and horns, sharing the full beam power between the four.
This system will be adopted here.

\subsection{The Target}

Each of the four targets will be hit by a 1.25~MW proton beam to produce the pions needed for the neutrino beam production.
Following the EURO$\nu$ studies, a packed bed of titanium spheres cooled with cold helium gas has become the baseline target design for a Super Beam based on a 2--5~GeV proton beam with a power of up to $\sim$1~MW per target.
Classical monolithic solid targets are almost impossible for this application because of the difficulty to provide efficient cooling.

There is a possible risk that the pulsed beam may generate vibrations of the spheres, which could be transmitted to the packed bed container and beam windows and cause degradation of the spheres where they are in contact with each other.
This problem is proposed to be studied in beam tests using the HiRadMat~\cite{Gaillard:2011zb, hiradmat-url} high intensity proton irradiation facility at CERN.

A Laser Doppler Vibrometer (LDV) will be used to measure the vibrations in the target container wall with the aim to compare locations where the spheres are in contact with the wall and where they are not.
The LDV will also be used to directly study individual spheres as the intense proton pulse interacts with them.
The titanium spheres will be contained in an open trough so that a fast camera can be used to view an open surface of the packed bed and detect any movement of the spheres caused by stress induced by the beam.
A similar test experiment using tungsten powder has already been carried out with promising results~\cite{powder-ipac}.
The continued studies will include computer simulations of the vibration of the spheres and their possible degradation at the contact points as well as further development of instrumentation.

The packed bed concept has been studied using Computation Fluid Dynamics (CFD) software tools.
However, the flow regime is complex and it is necessary to carry out prototype measurements in order to test the concept and gain confidence in the design.
An experimental program is planned using an induction heater power supply to generate heating of the individual spheres.
A short but representative length of the target proposed will be produced for the tests.
A well instrumented helium flow loop will be constructed to cool this target test piece.
The individual target spheres and container walls will be instrumented in order to measure both the effect of the heating and the cooling efficacy and make comparisons with the CFD model.

\subsection{The Horn and its Power Supply}

The hadron collector, also called horn, is used to focus in the forward direction the charged pions produced in the proton--target collisions.
As the neutrinos created in the decay of the pions tend to go in the same direction as the pions, a reasonably focused neutrino beam is thereby produced.
For the special case in which the target is placed inside the horn, because of the relatively low energy protons, to maximize the pion collection efficiency, the running conditions of the horn are particularly difficult.

The high pulse rate of 14~Hz, three times that in other current projects, and the higher level of radiation produced by the very high intensity of the proton beam, ten times more intense than what present proton drivers used to produce neutrino beams are providing, also represent a considerable difficulty.
These conditions could significantly reduce the system lifetime, requiring particular studies and tests to be performed.

The horn current pulse generator is a sizable item which must be placed as close as possible to the four horns to minimize the power dissipation in strip lines conducting the current from the pulse generator to the horns.
In the case of four accumulator rings and four target--horn assemblies and a single pulse generator, it is important to avoid having to pulse the horns simultaneously.
The beams in the four accumulator rings should therefore be ejected in sequence, after their simultaneous injection, with an equal time spacing of $17.9$~ms ((1/14~Hz)/4), implying that the current pulse generator needs to be operated at a frequency of 56~Hz.

A pulsed power supply able of providing the very high current (350~kA) to be circulated inside the horn at the required pulse rate has not been produced so far.
A first design of such a power supply supposed to be operated at 50~Hz has been produced by the EURO$\nu$ studies demonstrating on paper that the construction is feasible~\cite{Baussan:2013mua}.

As part of this project it is proposed the construction of a pulsed power supply prototype of one of the eight identical modules of the configuration proposed by EURO$\nu$.
This will allow tests of the characteristics of the system like power dissipation, current recovery system and lifetime.
The electrical properties of the horn are playing an important role for the requirements of the power supply.
It is also proposed to construct a horn prototype with a shape optimized for the ESS proton beam, which will enable tests of the horn when pulsed from the power supply, in particular studying the vibration levels, which limit the horn lifetime, and the efficiency of the horn cooling system.

\begin{table}[htdp]
\caption{Number of neutrinos per~m$^2$ crossing a surface placed on--axis at a distance of 100~km from the target station  during 200 days for 2.0~GeV protons and positive and negative horn current polarities.}
\begin{center}
\begin{tabular}{|c|c|c|c|c|}\hline
 & \multicolumn{2}{c|}{positive} & \multicolumn{2}{c|}{negative} \\
\cline{2-5} & $N_\nu\ (\times 10^{10})$/m$^2$ & \% &$N_\nu\ (\times 10^{10})$/m$^2$ & \% \\
\hline $\nu_\mu$           & 396 & 97.9     & 11   & 1.6 \\
\hline $\bar\nu_\mu$ & 6.6   & 1.6     & 206 & 94.5 \\
\hline $\nu_e$               & 1.9   & 0.5     & 0.04   & 0.01 \\
\hline $\bar\nu_e$     & 0.02   & 0.005 & 1.1   & 0.5 \\
\hline
\end{tabular}
\end{center}
\label{fluence}
\end{table}

\begin{figure}[hbt]
\begin{center}
\includegraphics[width=1.0\textwidth]{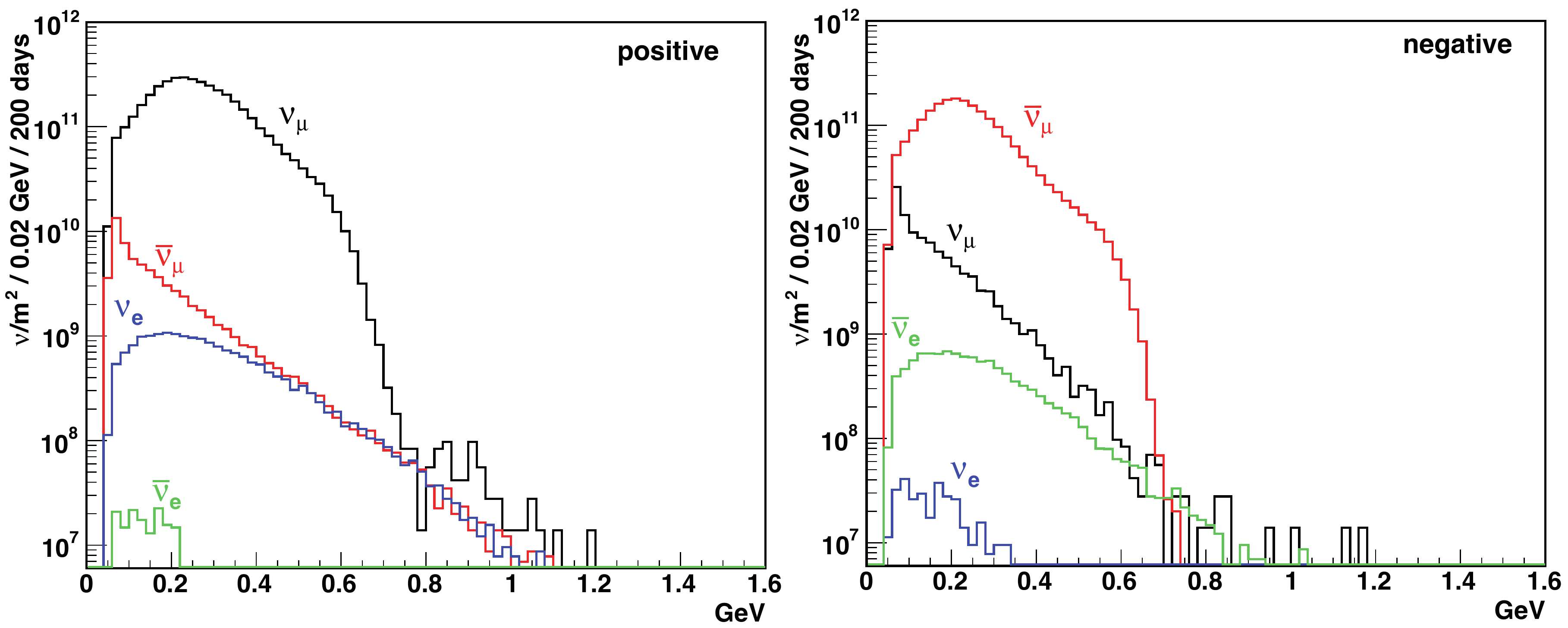}
\end{center}
\caption{Neutrino fluence as a function of energy at a distance of 100~km on--axis from the target station, for 2.0~GeV protons and positive (left) and negative (right) horn current polarities, respectively.}
\label{nuspectrum}
\end{figure}

\begin{figure}[hbt]
\begin{center}
\includegraphics[width=0.7\textwidth]{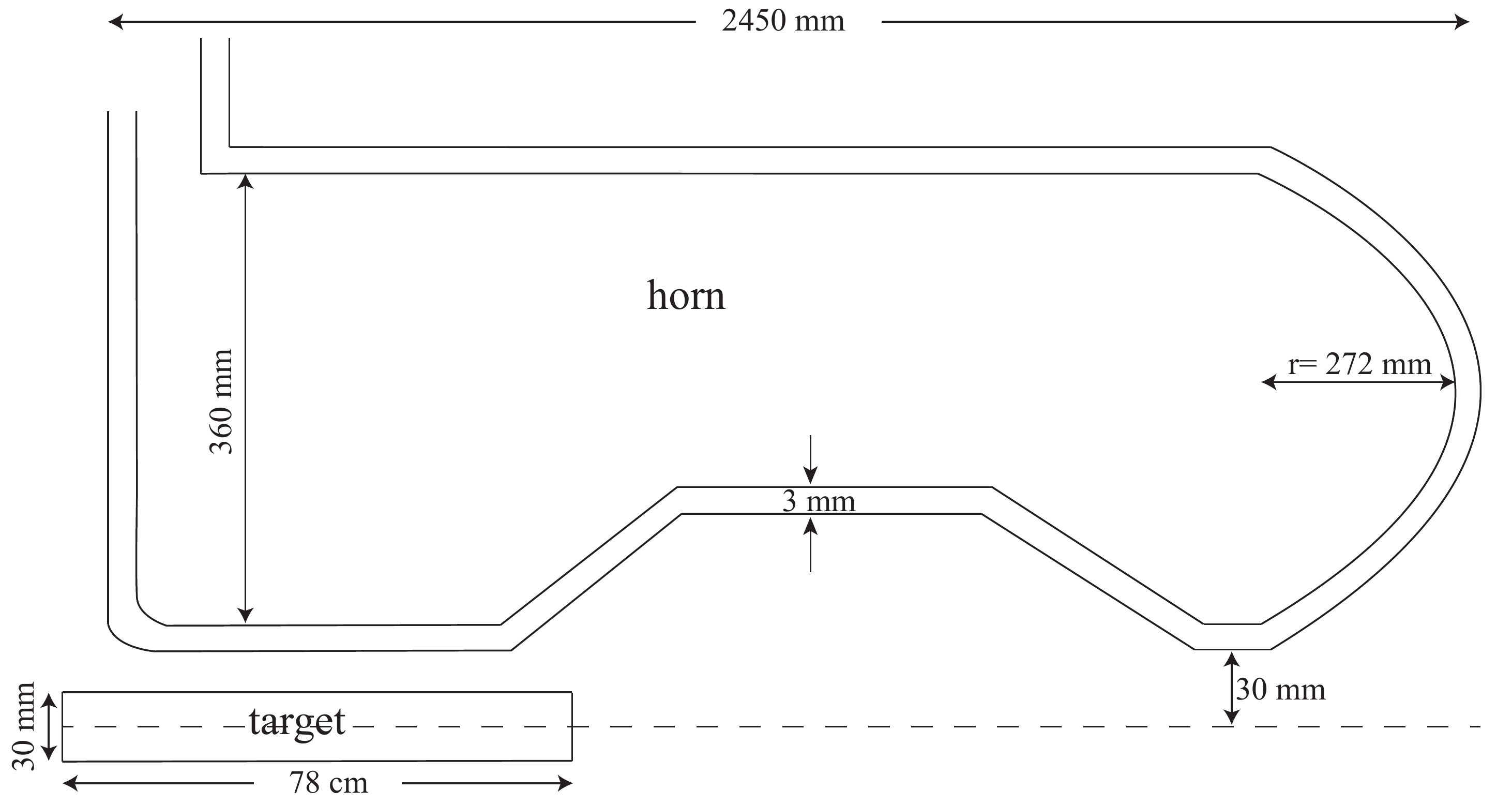}
\end{center}
\caption{Horn and target geometry used for these studies (not to scale).}
\label{nikosxx}
\end{figure}

Fig.~\ref{nuspectrum} displays the total number of neutrinos per~m$^2$ crossing a surface placed on--axis at a distance of 100~km from the target station  during 200 days ($1.73\times 10^7$~s).
Table~\ref{fluence} presents the number of neutrinos corresponding to Fig.~\ref{nuspectrum}.
The total contamination for positive (negative) polarity coming from $\bar\nu_\mu$ ($\nu_\mu$), $\bar\nu_e$ and $\nu_e$ is 2.1\% (5.5\%).
This has been obtained using the target and horn geometry shown in Fig.~\ref{nikosxx}.

\begin{figure}[hbt]
\begin{center}
\includegraphics[width=0.5\textwidth]{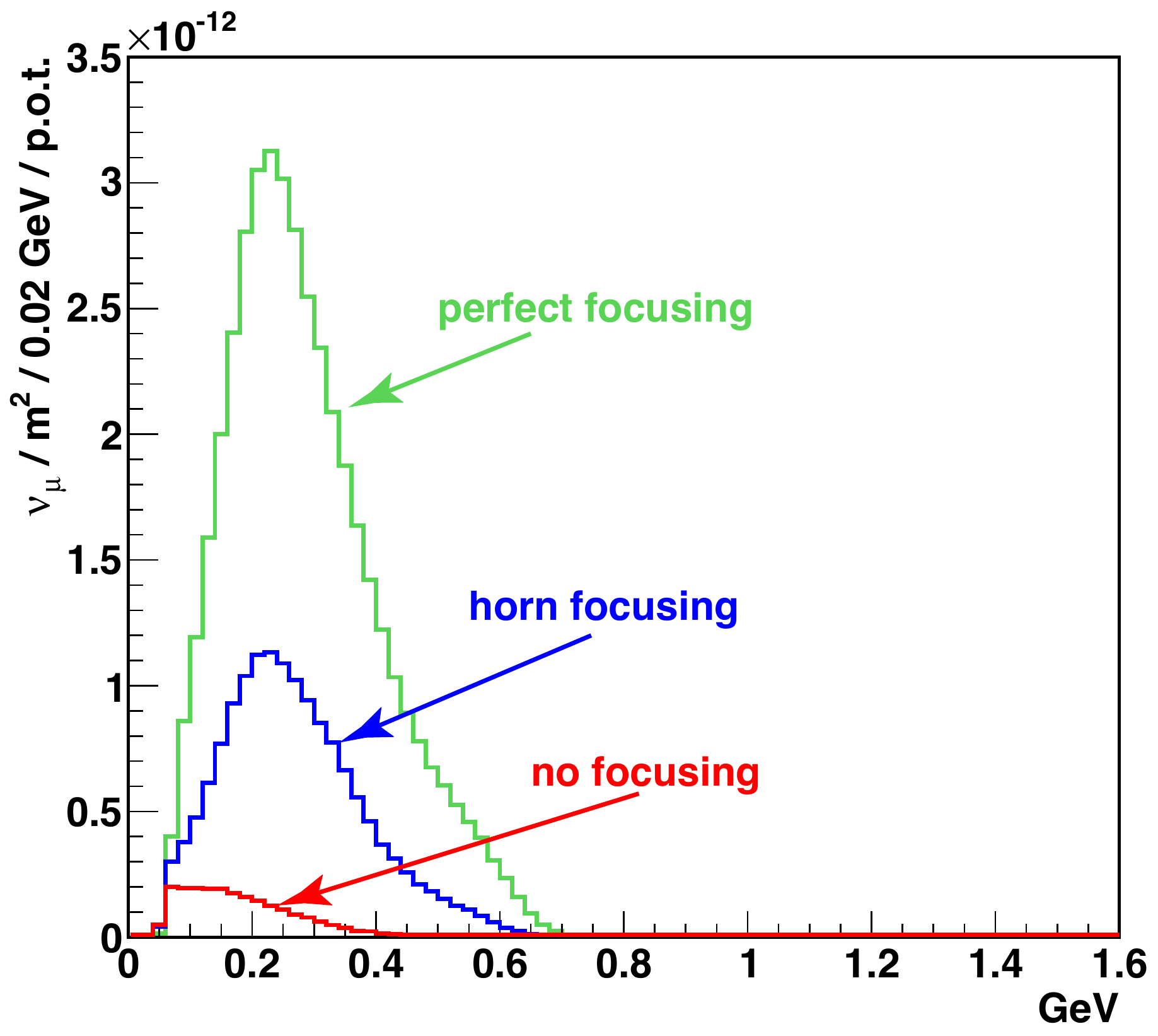}
\end{center}
\caption{Neutrino energy distribution without focusing (red, no current in the horn), with focusing (blue) and for a perfect focusing (green), respectively, for 2.0~GeV protons and positive current polarity.}
\label{merit}
\end{figure}

Fig.~\ref{merit} presents the neutrino energy distribution without focusing (no current in the horn), with focusing using the horn and for a perfect focusing  (i.e. all produced pions are directed towards the detector).
The horn focusing allows to enhance the number of neutrinos directed towards the detector by a factor 7.4.

\section{Underground Detector Site}
\label{detector}

In the search for a suitable site for the large underground Water Cherenkov detector some preliminary investigations have been made of the Northern Garpenberg mine at 540~km NNW of the ESS site in Lund.
The construction of the ore hoist shaft of this mine and the nearby decline (descending transport tunnel) started in the 1960s.
The current shaft depth of 830~m was reached in 1994 and the depth of the decline (of cross section $5\times6$~m$^2$), which was 1000~m in 1998, has later been extended to 1230~m.
In 2012 300Õ000~tons (=110Õ000~m$^3$) of ore was transported with trucks on the decline up to the shaft hoist at 830~m depth and hoisted up to the ground level.
The hoist, shaft and head--frame (hoist surface tower) will no longer be used as from end of 2014.
To preserve them will require their maintenance.

The dominating limestone and dolomite rock in the 2~km broad and 10~km long syncline of the Garpenberg mine, oriented in the north--east direction, is not suitable for the excavation of the very large MEMPHYS type of detector caverns.
However, the syncline is surrounded by granite which is of sufficient strength and homogeneity to allow the safe construction of large caverns. 
Bore holes have been drilled from the mine at 880~m depth out to the granite zone and the mean value of the uniaxial compressive strength of the granite bore core samples collected has been measured to be 206~MPa.
The rock stress at 880~m depth has previously been measured.
The main component of the stress is horizontal and directed to the north--west.
Its magnitude is about 40~MPa.
On our request Garpen Mining Consultant Inc has evaluated the geological and geotechnical data of the surroundings of the Northen Garpenberg mine.
The Consultant concludes that the data are similar to those reported for the planned Pyh\"{a}salmi~\cite{laguna-url} neutrino detector sites and that the possibilities are good for the construction and installation of a MEMPHYS size detector at the the Northen Garpenberg mine. 

In order to make a full exploration of the conditions at the Garpenberg mine a series of bore holes should be drilled from the bottom of the decline at 1230~m depth in different directions in a sector from west to north.
As already mentioned, recently ESS has started a study of a design of the linac for a proton energy of  2.0~GeV, instead of  as in the original base line design 2.5~GeV.
Our preliminary simulation results indicate that the performance for CP violation discovery with a 2.0~GeV proton energy beam is somewhat better with the base line 360~km of the Zinkgruvan mine, situated at the northern tip of lake V\"attern in Sweden,  than with the 540~km base line of the Garpenberg mine.
A possible option would be to accelerate the beam in the accumulator ring from 2.0~GeV to 2.5~GeV for which energy the CP performance with these two baselines is about equivalent and somewhat better that with 2.0~GeV proton energy and the 360~km baseline (see Fig.~\ref{fractions3s5s}).
However, in view of the possibility that the proton energy will be 2.0~GeV, we intend to make preliminary studies of the conditions for detector installation and operation also at the Zinkgruvan mine.
One significant difference is that there is no plan at Zinkgruvan, like at Garpeneberg, to free a shaft and ore hoist that could be used to transport the detector--hall excavation rock masses.
At Zinkgruvan a second shaft with a hoist would therefore have to be constructed for that purpose.

Once the proton beam energy has been definitely decided and the optimal baseline has been obtained from refined simulations, the final choice between these two mines will be made on the basis of the calculated optimal baseline, the geological parameters and the existing mining infrastructures.
The selected mine will then be studied in further detail collecting geological and rock mechanics information at potential detector locations, situated  at 1000~m depth (~3000~m water equivalent) and at least 500~m from locations with active mining operations, by making core drillings, core logging, rock strength testing and rock stress measurements of the surrounding rock. 

Once a suitable location for the neutrino detector underground halls, which should have a total volume of 6$\times 10^5$~m$^3$, has been determined, a design of the geometry and construction methods for the underground halls will be made based on the measured strength and stress parameters of the rock.
The technical part of this task will be subcontracted to a rock engineering company.
The task will include an estimation of the cost of the underground hall excavation and reinforcement work.

As detector for the appearing electron nutrinos we propose to use a large water tank Cherenkov detector.
Although the MEMPHYS detector has been extensively studied by LAGUNA, readjustments of the shape of the detector volumes will be made according to the rock quality at the chosen location. The effect of these readjustments on the detector efficiency will be evaluated and integrate in the whole physics performance evaluation of the project.

\section{Physics Performance}

An important parameter to be determined is the optimal neutrino beam baseline, given the parameters of the achievable neutrino beam.
The simulation software developed by the EURO$\nu$ project has already been used to make first evaluations of the potential for leptonic CP violation and neutrino mass hierarchy discoveries.
In particular, the fraction of the full CP violation phase range within which CP violation and neutrino mass hierarchy can be discovered, at different baselines, has been computed~\cite{1212.5048}.

According to these first evaluations, for which  5\% systematic error on the signal and 10\% systematic error on the background were assumed, leptonic CP violation could be discovered at 5~$\sigma$ confidence level within at least 50\% of the CP phase range for baselines in the range 300-550~km with an optimum of about 58\% of the phase range at a baseline of about 420~km, already an excellent physics performance.
According to the same first evaluations, the neutrino mass hierarchy can be determined at more than 3~$\sigma$ confidence level for baselines in the range 300--500~km depending on the proton beam energy.
In addition, inclusion of data for atmospheric neutrino oscillations in the mass hierarchy determination will certainly improve the physics reach of this project.
For these first evaluations, the target, horn and decay tunnel parameters, had not been optimized for the ESS proton driver energy of 2.0~GeV.

The study and optimization of the physics performance will be continued using the new input parameters resulting from the Super Beam component simulations and prototype tests and improvements of the physics simulation software.
After all optimizations of the simulation analyses the overall expected performance of ESS$\nu$SB will be evaluated.

Most of the neutrinos derived from the 2.0~GeV proton beam will have energies in the range 200--500~MeV.
For such comparatively low energies the rate of inelastic events is limited, implying that it is sufficient to measure only the outgoing charged lepton in the event.
Furthermore, the neutrino cross--section is lower than at higher energies, leading to the requirement of a comparatively large target volume.

Till now the foreseen ESS linac proton energy has been and still is 2.5~GeV with the possibility of a future upgrade to at least 3.0~GeV.
Currently an initial running period starting 2022 with 2.0~GeV protons is being discussed.
In view of this we have simulated the physics performance of ESS$\nu$SB for the  proton energies 2.0~GeV, 2.5~GeV and 3.0~GeV.
For technical limitations, the proton energy inducing the most severe conditions is considered (radiations, space charge effects, cooling etc.).

\begin{figure}[hbt]
\begin{center}
\setlength\fboxsep{0.6pt}
\setlength\fboxrule{0.6pt}
\mbox{
 \includegraphics[width=0.8\textwidth]{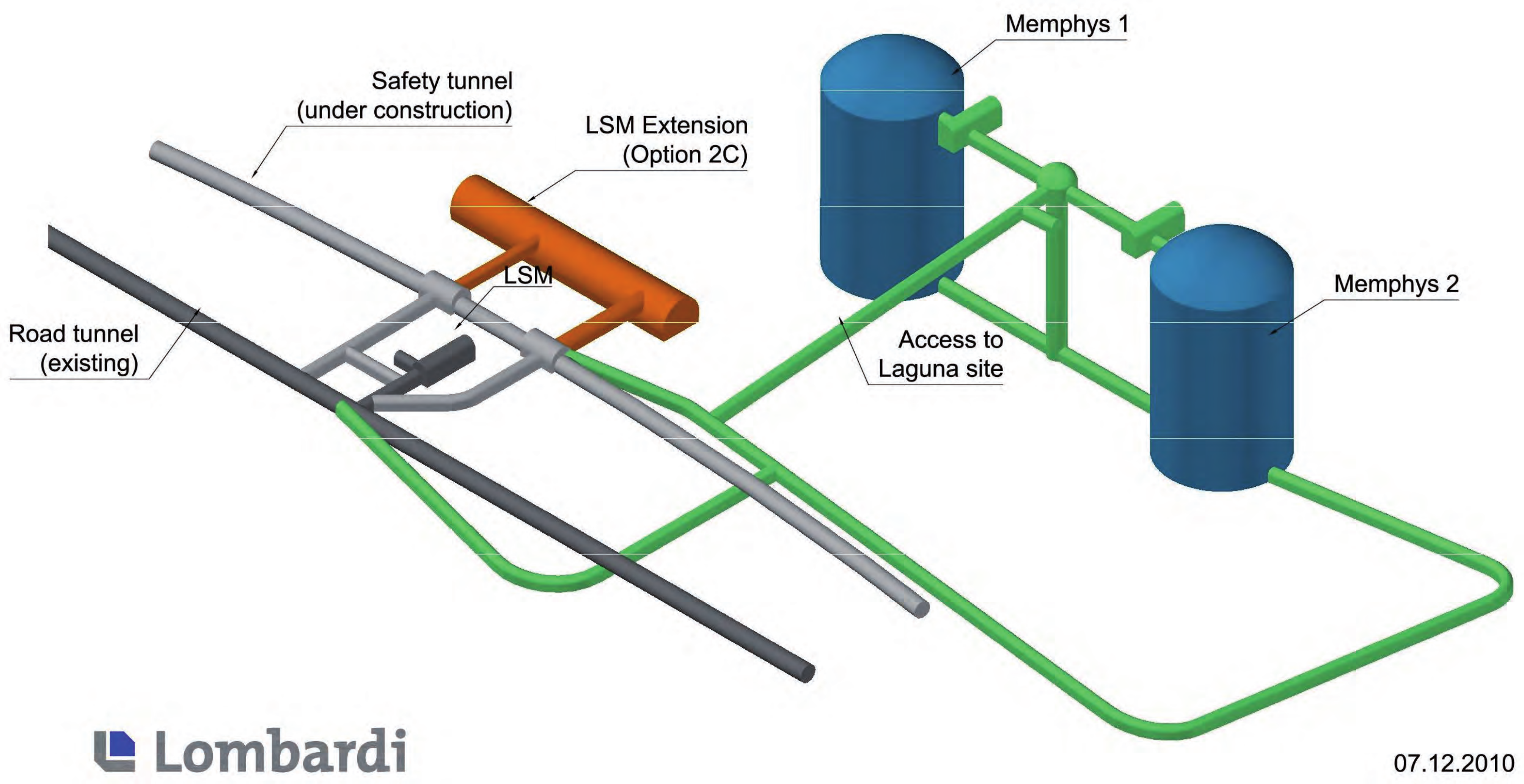}}
\end{center}
\caption{The MEMPHYS Water Cherenkov detector (courtesy of Lombardi Engineering S.A.) as could be installed in the Fr\'ejus tunnel. The total fiducial mass is 500~kton consisting of two cylindrical modules of 103~m high and 65~m diameter and with 2$\times$100000 12'' photomultipliers mounted on the walls, providing 30\% geometrical coverage.}
\label{memphys}
\end{figure}

For ESS$\nu$SB the neutrino detector will be a Water Cherenkov detector with a fiducial volume of 500~kt of the same type as the MEMPHYS detector \cite{hep-ex/0607026} (Fig.~\ref{memphys}) planned for the Fr\'ejus site (with the CERN SPL as proton source for the neutrino beam).
In comparison, a liquid scintillator or a liquid argon detector (technologically challenging and significantly more difficult to operate), are more expensive by unit volume but are needed for the reconstruction of the inelastic events that are more frequently produced at significantly higher energies than those foreseen at the ESS.

For the calculations presented below we have used the simulation tool GLoBES\footnote{A modified version of GLoBES, which includes a near detector and correlations between systematic errors affecting different channels, has been used for all the results presented in this work. For details on the $\chi^2$ and systematics implementation, see~\cite{1209.5973}.}\cite{Huber:2004ka, Huber:2007ji} elaborated for the MEMPHYS detector with the CERN SPL as neutrino source.
For the neutrino flux calculation, the proton energy has been set to 2.0~GeV and the protons on the target to $2.7\times10^{23}$ per year.
For any other energy a constant proton power 5~MW has been assumed.


The large Water Cherenkov detector can also be used, concurrently, for astroparticle physics and proton lifetime measurements.
During their explosion, supernovae emit a considerable amount of neutrinos.
The modelling of the explosion mechanism has been studied for more than 30 years~\cite{1112.1913}.
The detection of the emitted neutrinos could help to understand this mechanism.
Moreover, a galactic supernova explosion will be the occasion to explore the neutrino properties at distance scales of the order of $10^{17}$~km and time scales of about $10^5$ years.
It is believed that core--collapse supernovae have been occurring throughout the universe since the formation of stars.
There should be a diffuse background existing of neutrinos originating from these cataclysmic cosmological events.
The detection of these relic neutrinos~\cite{Bays:2011si} would give information about the history of star formation.
The proposed neutrino detector for the ESS neutrino Super Beam facility will have a detection threshold lower than 10~MeV and could detect neutrinos coming from supernova explosions provided that the detector is located sufficiently deeply underground to be protected from cosmic rays.
This is one of the reasons why the proposed detector location will be 1000~m underground.
The large fiducial mass of the detector of $\sim$500~kt will make possible the detection of about $5\times10^4$ events from galactic stellar collapses during ten years of operation.

Proton decay is not allowed by the Standard Model.
On the other hand, Grand Unified Theories predict proton decay.
Its discovery would reveal the existence of a more fundamental theory beyond the Standard Model.
The present lifetime lower limit for the decay $p\rightarrow \pi^0e^+$ is $5\times10^{33}$~yrs.
This limit was set by Super--Kamiokande experiment~\cite{Nishino:2012ipa} employing the same Water Cherenkov detector technique but with a detector volume 20 times smaller than that proposed for the Super Beam.
The Super Beam detector would be able to reach a proton lifetime limit of $10^{35}$~yrs in ten years running.

\subsection{Optimization of the Baseline}

In order to evaluate numerically the discovery potential of leptonic CP violation for different baselines $L$ and for different values of with the leptonic Dirac CP--violating phase~$\delta_{CP}$, GLoBES has been used to simulate the neutrino oscillations and the detection of the neutrinos in the MEMPHYS detector, varying $L$.
The parameter values used in the GLoBES calculation are:~$\Delta $m$^2_{21}$ = 7.5$\times 10^{-5}$ eV$^2$,~$\Delta $m$^2_{31}$ = 2.47$\times 10^{-3}$ eV$^2$, $\theta_{21}$ = 0.58,  $\theta_{13}$ = 0.15 and $\theta_{23}$ = 0.70.
These values roughly correspond to the global fit of~\cite{1209.3023}.
The neutrino mass hierarchy is not assumed to be known.

These parameters are included in the fit assuming a prior knowledge with an accuracy of 3\% for $\theta_{12}$, 0.02 for $\sin^22\theta_{23}$, 4\% for~$\Delta $m$^2_{31}$ and 3\% for~$\Delta $m$^2_{12}$ at 1~$\sigma$ level.
The error in  $\sin^22\theta_{13}$ has been set to 0.005, which corresponds to the expected precision limit of the Daya Bay experiment.
All other systematics are described in~\cite{1209.5973} (Table~2 ``default'' case).

\begin{figure}[htdp]
\begin{center}
\includegraphics[width=1.0\textwidth]{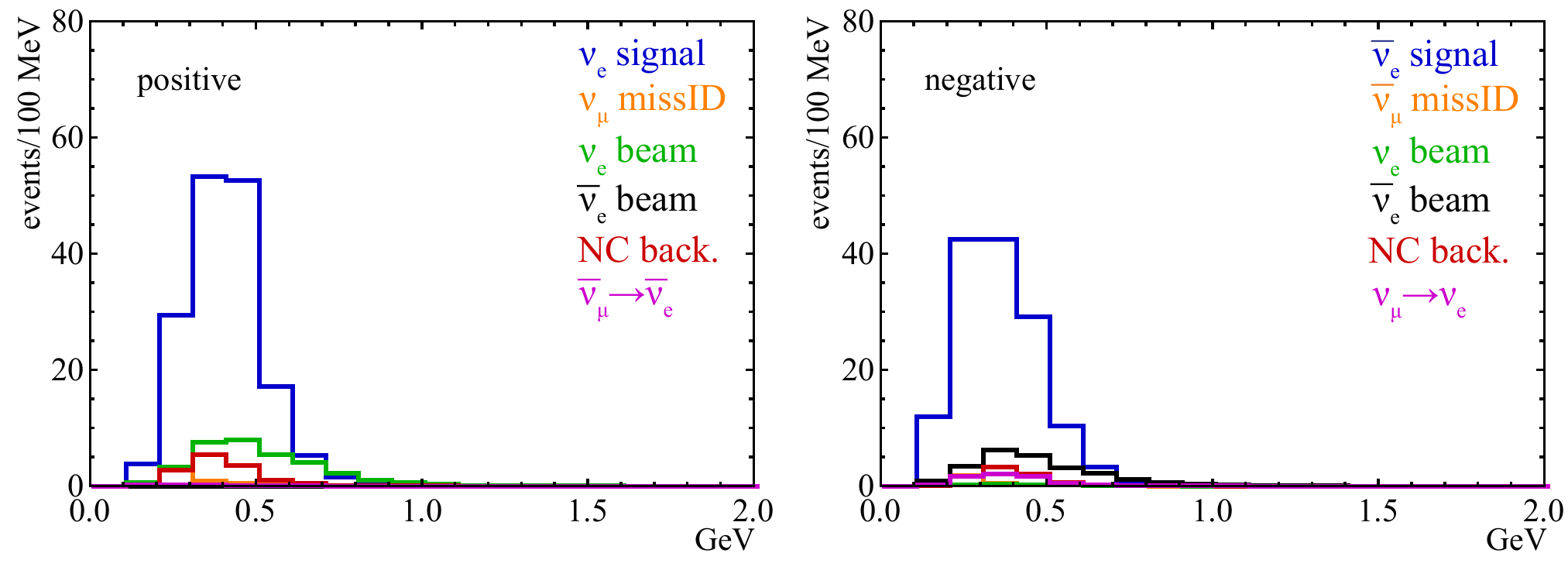}
\end{center}
\caption{Energy distributions of the detected electron neutrinos (positive) and anti--neutrinos (negative) including background contribution as reconstructed by MEMPHYS detector for two years of neutrino running (left) plus eight years of antineutrino running (right) and a baseline of 540~km (2.0~GeV protons, $\delta_{CP}=0$).}
\label{spectrum}
\end{figure}

The detector performance in terms of efficiencies, backgrounds and event migration follows the analysis of~\cite{1206.6665}.
The data collection period assumed is two years of neutrino running plus eight years of antineutrino running in order to detect in the far detector about the same total number of the two kinds of neutrinos.
Fig.~\ref{spectrum} shows the electron and anti--electron neutrino spectra as detected by the far detector for a baseline of 540~km, which is near the second oscillation maximum ($\delta_{CP}=0$).
Also shown are the contributions from various background sources.
The neutrino mean energy of the detected electron neutrinos is approximately 350~MeV with a large FWHM of about 300~MeV and a tail towards higher energies.
Table~\ref{neutrinos} gives the number of all neutrino sources.

\begin{table}[hbt]
\caption{Number of neutrinos for two plus eight years running with neutrinos and anti--neutrinos respectively (Fig.~\ref{spectrum}).}
\begin{center}
\begin{tabular}{|r|c|c|c|c|c|c|}
\hline  experiment & $\nu_e$ ($\bar\nu_e$) & $\nu_\mu$ ($\bar\nu_\mu$) & $\nu_e $   & $\bar\nu_e$ & NC     & $\bar\nu_\mu(\nu_\mu)\rightarrow\bar\nu_e(\nu_e)$  \\
 configuration         & signal                                 & miss--ID                                      & beam   & beam                & back. &   \\
\hline 360~km positive       & 303.3                               & 10.7           & 70.8                          & 0.08          &   29.2    & 1.4         \\
                           negative     & 246.1                                & 6.1           & 2.4                            & 50.6           &     17.4   & 13.3          \\
\hline 540~km positive      & 196.7                                & 4.6           & 33.3                           & 0.04           &      13.7 & 0.9        \\
                          negative     & 162.9                               & 2.8          & 1.1                             & 23.5           &     8.2      & 7.8         \\
\hline
\end{tabular}
\end{center}
\label{neutrinos}
\end{table}

\subsubsection{CP Violation}

The GLoBES code has been used to calculate the number of detected electron neutrinos and anti--neutrinos for values of $\delta_{CP}$ ranging from -180 to 180 degrees and for different baselines $L$ and proton energies.
For fixed baseline $L$ and for each value of $\delta_{CP}$ the number of events was generated and the $\chi^2$ was computed for $\delta_{CP} = 0$ and $\delta_{CP}= \pi$ in the two neutrino mass hierarchy orderings, marginalized over all other oscillation parameters within their corresponding priors.
The smallest of these four values (the best fit to a CP conserving value) was recorded.
The square root of this number provides the significance in terms of standard deviations $\sigma$ with which CP violation could be discovered for 2.0~GeV protons.
This calculation was carried out for a series of different baselines varying from 200~km to 800~km.

For all performance results reported below the systematic errors recommended in~\cite{1209.5973} have been used and implemented for the {\it snowmass} 2013 process~\cite{snowmassref-url} neutrino studies.
These systematic errors are more conservative than those assumed in the first performance evaluation (5\%/10\% systematic error for the signal/background) reported in~\cite{1212.5048}.
The choice to present the more conservative results obtained under the systematic error assumptions of~\cite{1209.5973} has been done in the interest of enabling a comparison of these results with those of other experimental proposals
using the same performance evaluation tools.
These calculations also include the simulation of a near detector as specified in~\cite{1209.5973}.
Furthermore, for the results presented below, the neutrino target parameters (horn current, target position in the horn and the length and diameter of the decay tunnel), optimized for the 4.5~GeV proton energy of SPL, have so far only partially been re--optimized for the ESS 2.0~GeV energy.


\begin{figure}[hbt]
\begin{center}
  \includegraphics[width=0.6\textwidth]{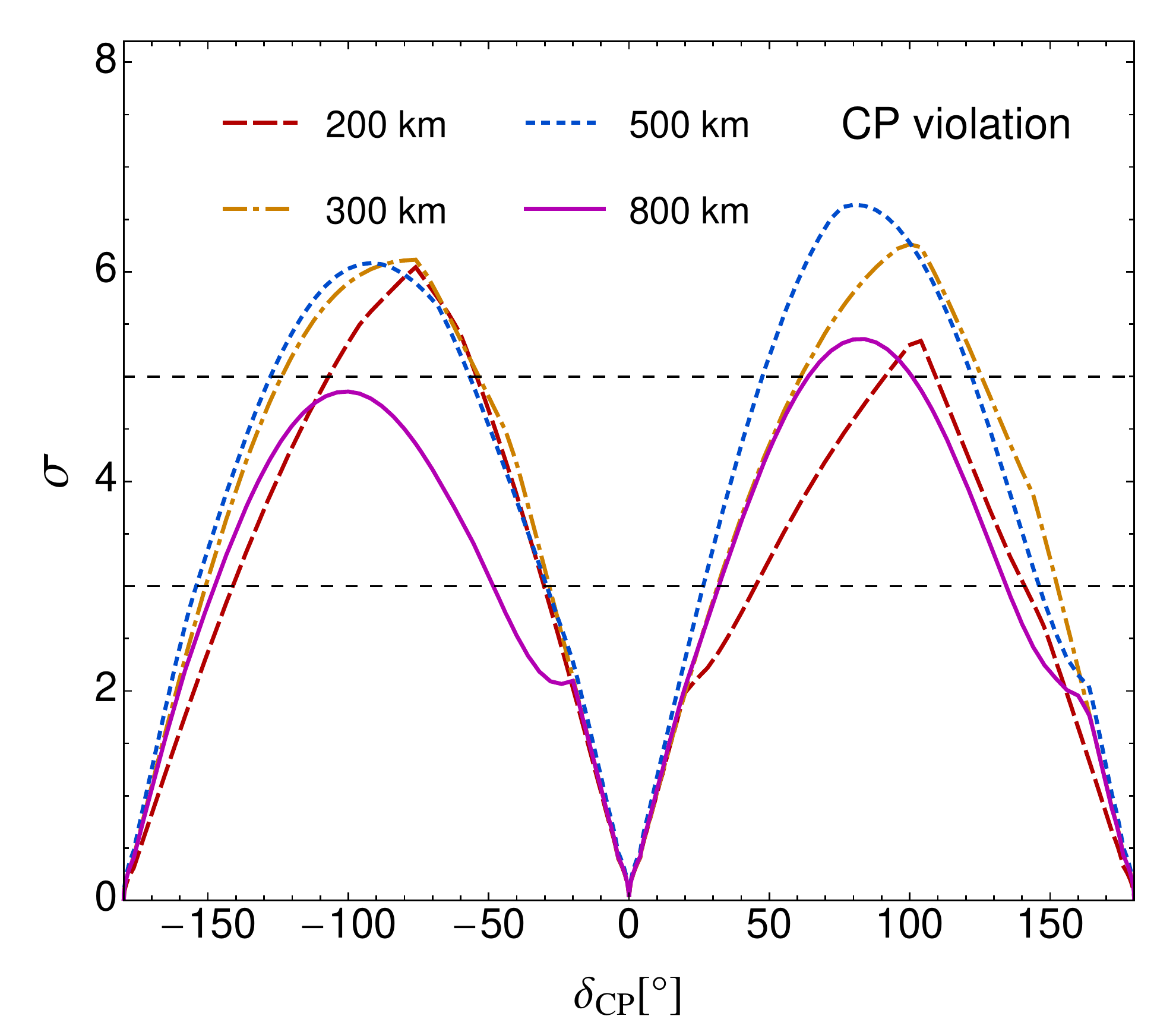}
\end{center}
\caption{The significance in terms of number of standard deviations $\sigma$ with which CP violation could be discovered for~$\delta_{CP}$-values from -180$^{\circ}$ to 180$^{\circ}$ and for different baselines $L$ (2.0~GeV protons).}
\label{deltacp}
\end{figure}

In Fig.~\ref{deltacp}  is shown the significance in terms of number of standard deviations $\sigma$ with which CP violation could be discovered as a function of the value of~$\delta_{CP}$ from -180$^{\circ}$ to 180$^{\circ}$.
The different curves represent different distances $L$.
The two horizontal lines have been drawn at the significance levels 3~$\sigma$ and 5~$\sigma$.
It is seen that the best performance is obtained for $L\sim 400-500$~km.

\begin{figure}[hbt]
\begin{center}
\includegraphics[width=0.6\textwidth]{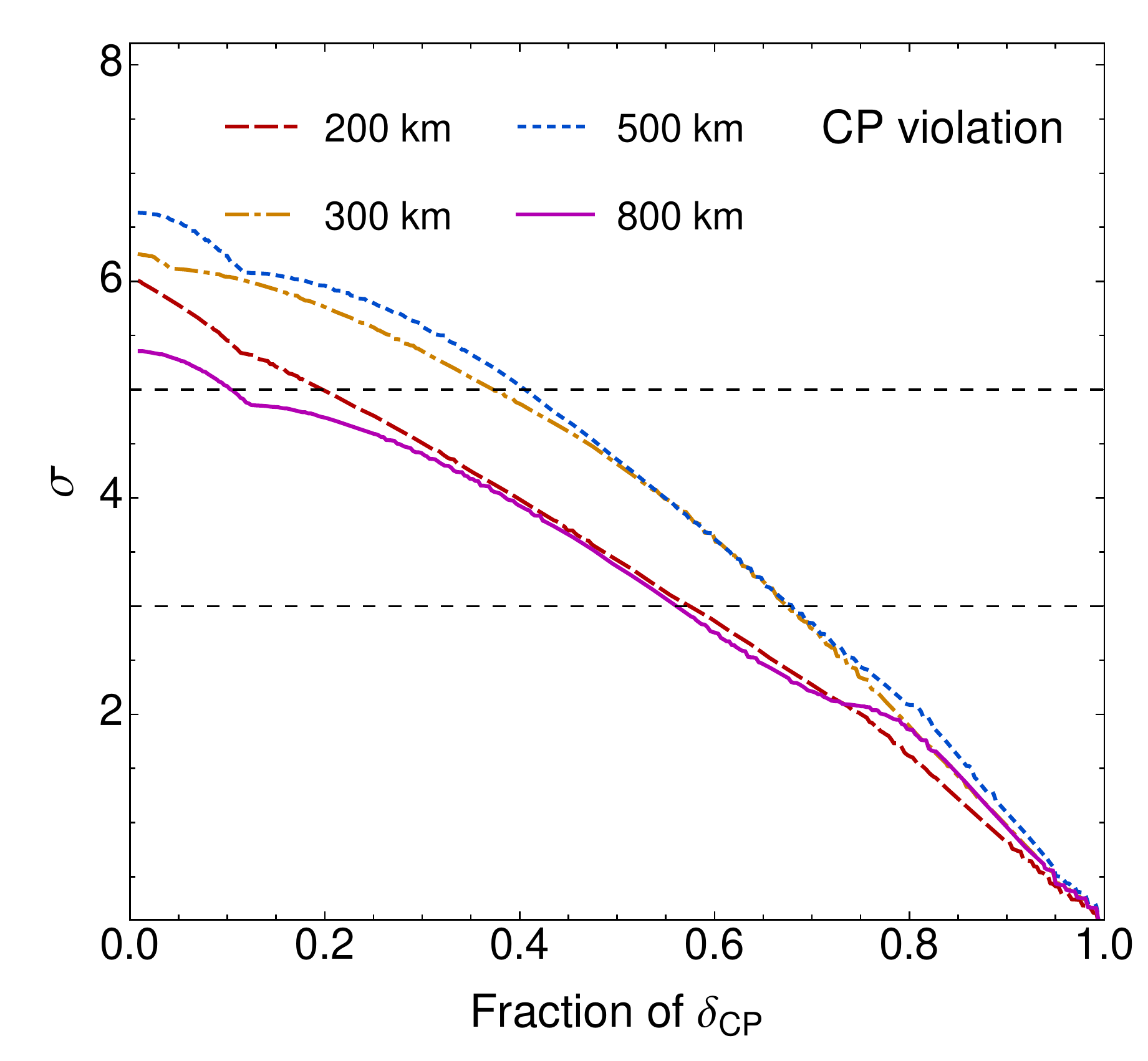}
\end{center}
\caption{The significance in terms of number of standard deviations $\sigma$ with which CP violation can be discovered as function of the fraction of the full~$\delta_{CP}$ range for different baselines $L$ (2.0~GeV protons).}
 \label{deltacpfrac}
\end{figure}

In Fig.~\ref{deltacpfrac} is shown the significance in terms of number of standard deviations $\sigma$ with which CP violation could be discovered as function of the fraction of the full~$\delta_{CP}$ range from -180$^{\circ}$ to 180$^{\circ}$ for which this discovery is possible.
As already noted above, the best performance is obtained for a baseline of the order of 300~km to 500~km where about 40\% of $\delta_{CP}$ range is covered with 5~$\sigma$ significance.

\begin{figure}[hbt]
\begin{center}
 \includegraphics[width=0.6\textwidth]{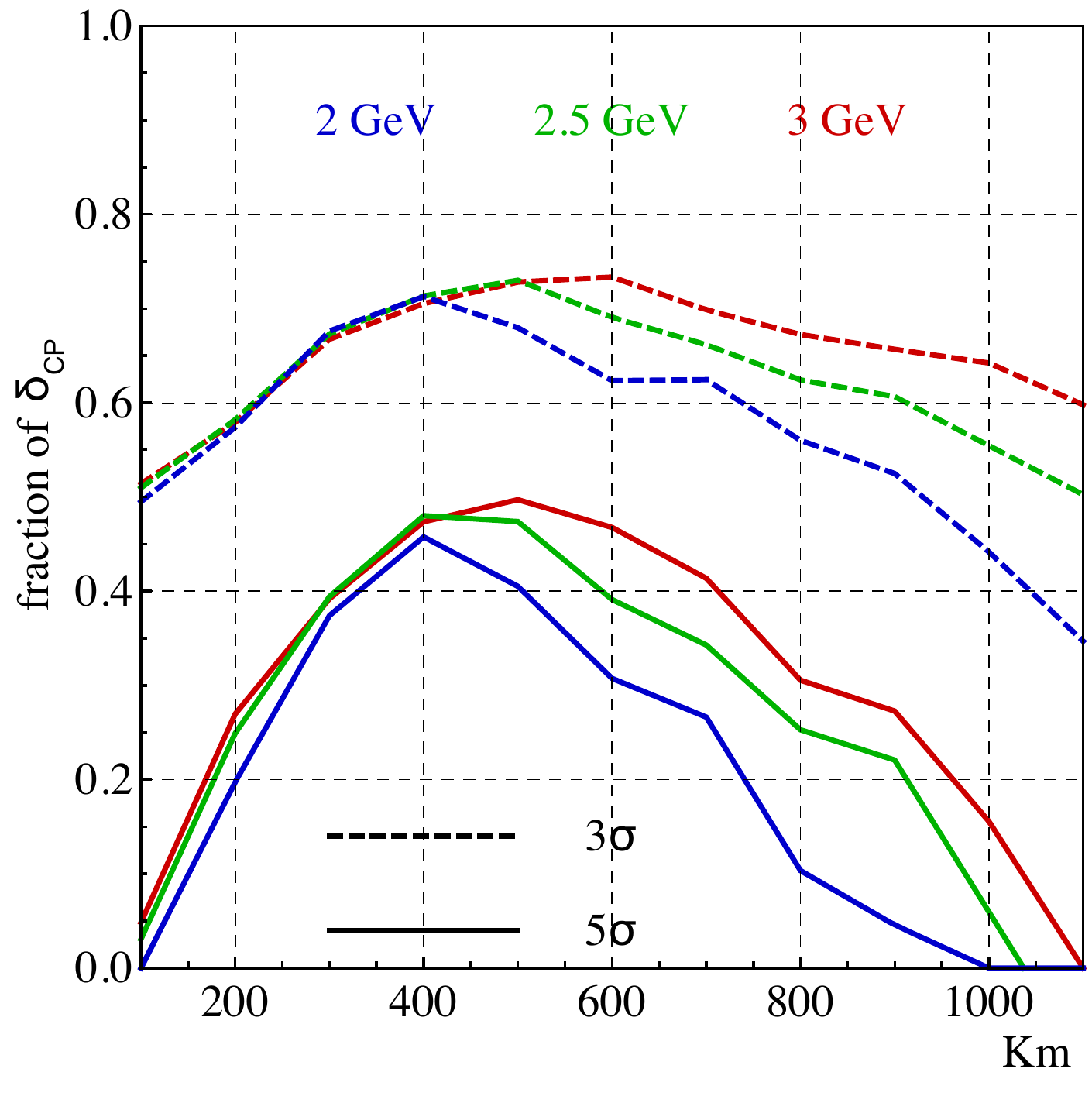}
\end{center}
\caption{The fraction of the full~$\delta_{CP}$ range for which CP violation could be discovered as function of the baseline. The lower (upper) curve is for CP violation discovery at 5~$\sigma$ (3~$\sigma$) significance.}
 \label{fractions3s5s}
\end{figure}

Fig.~\ref{fractions3s5s} presents the fraction of the full~$\delta_{CP}$ range (-180$^{\circ}$ to 180$^{\circ}$) within which CP violation can be discovered as function of the baseline in~km and for proton energies from 2.0~GeV to 3.0~GeV. 
According to the results of these calculations the fraction of the full $\delta_{CP}$ range within which CP violation can be discovered at 5~$\sigma$ (3~$\sigma$) significance is above 40\% (67\%) in the range of baselines from 300~km to 550~km and has the maximum value of 50\% (74\%) at around 500~km for 3.0~GeV.

\begin{figure}[hbt]
\begin{center}
 \includegraphics[width=0.49\textwidth]{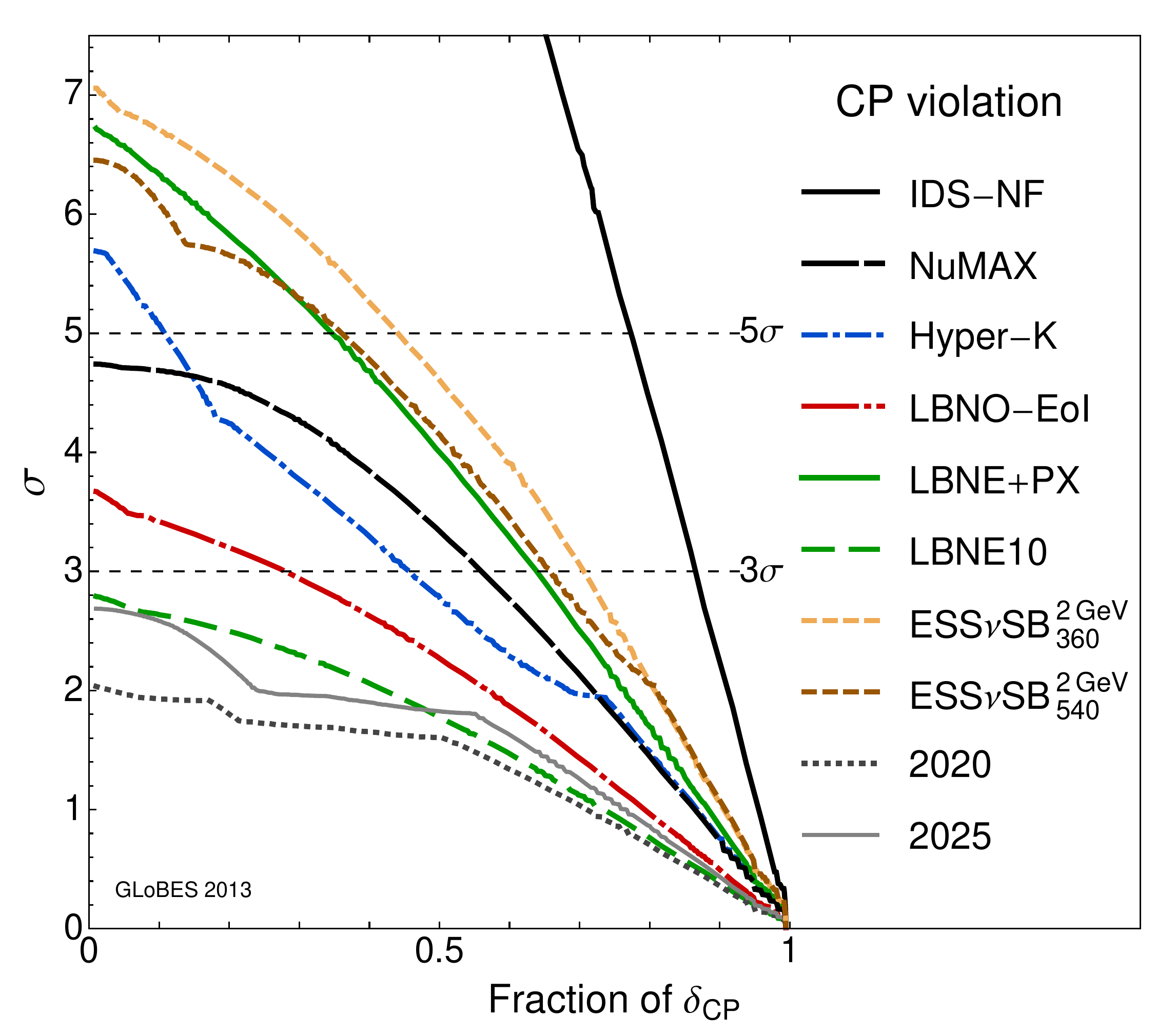}
 \includegraphics[width=0.49\textwidth]{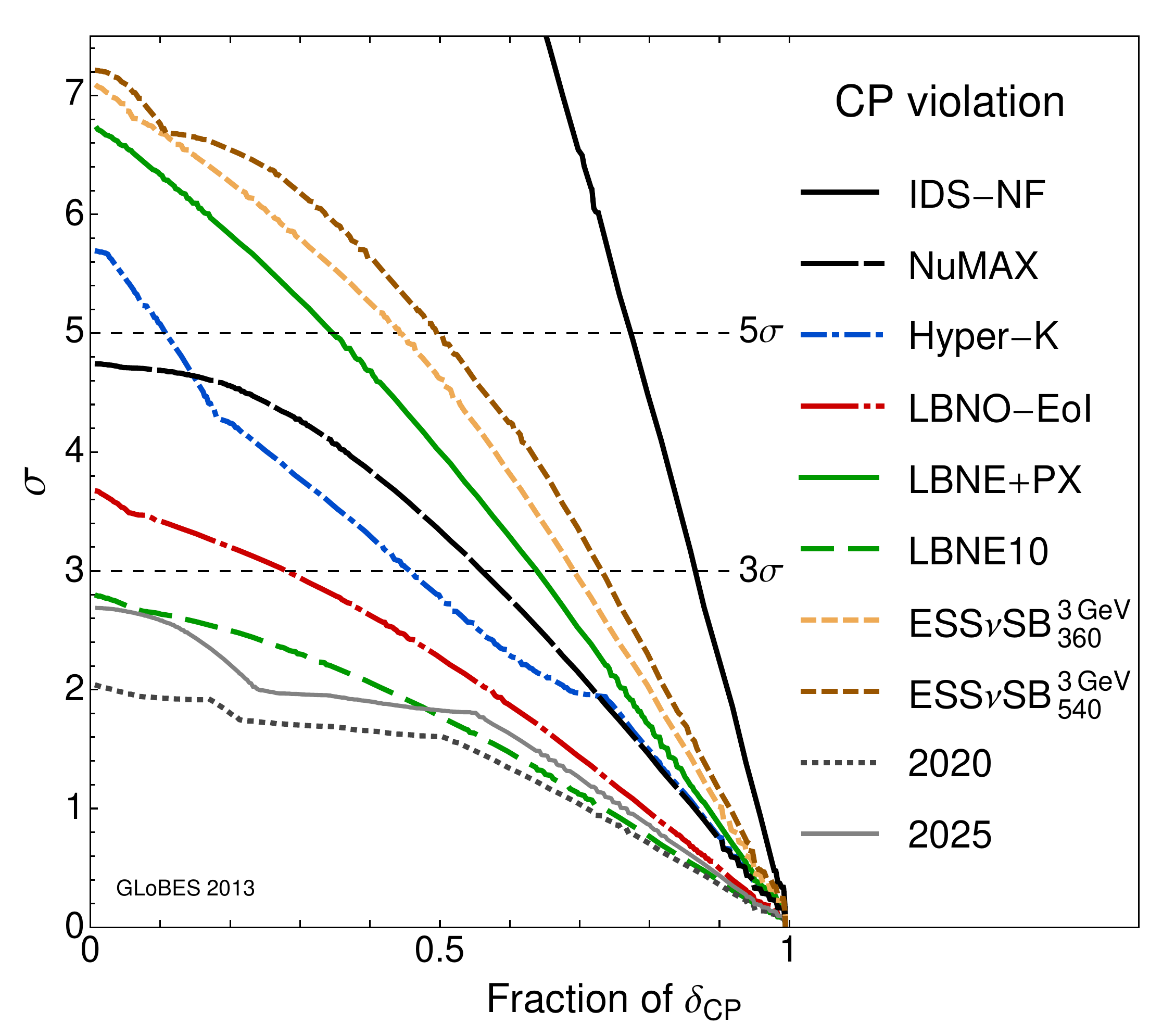}
\end{center}
\caption{The significance in terms of number of standard deviations $\sigma$ with which CP violation can be discovered as function of the fraction of the full~$\delta_{CP}$ range for different proposed experiments.
For ESS$\nu$SB the two baselines of 360~km and 540~km and two proton energies (2.0~GeV on left and 3.0~GeV on right) are shown. ``2020" considers 3+3 years of NOvA, and 5 years only for neutrinos in T2K (at its nominal luminosity, 0.75~MW); ``2025" considers 5+5 years of NOvA, and 5+5 years for T2K. The detector simulation details for T2K follow~\cite{0907.1896}, while for NOvA see~\cite{1209.0716, gina-comm}.}
 \label{compare1}
\end{figure}

Finally, Fig.~\ref{compare1} ({\it snowmass} 2013 process~\cite{snowmassref-url}), which is of the same kind as Fig.~\ref{deltacpfrac}, shows a comparison, for unknown mass hierarchy, of the ESS$\nu$SB performance for a baseline of 540~km and two proton energies (2.0~GeV and 3.0~GeV), with the performance of other proposed facilities.
 Only the much more advanced and costlier~\cite{cost-url} low energy Neutrino Factory (IDS-NF) would perform better than the ESS Neutrino Super Beam.
The main parameters used for all facilities are summarized in Table~\ref{experiments} while the considered systematic errors are those reported in~\cite{1209.5973} (for ESS$\nu$SB see SB in Table~2 ``default'' case).
As already said, the more optimistic systematic errors of signal/background of 5\%/10\% have been used in~\cite{1212.5048} for ESS$\nu$SB, where the CP violation coverage can go up to 59\% (78\%) at 5~$\sigma$ (3~$\sigma$).
 Efforts are currently done to find ways to reduce the systematic errors (and demonstrate that ``optimistic'' case of Table~2 in~~\cite{1209.5973} is reachable) using a high performance near detector and the possibility to measure the relevant electron neutrino cross--sections using this near detector and $\nu_e$ and $\bar\nu_e$ (contamination) contained in the ESS$\nu$SB neutrino beam (see Table~\ref{fluence}).
These cross-sections could also be measured by $\nu$STORM~\cite{1206.0294}.

\begin{table}[hbt]
\caption{Conditions under which Fig.~\ref{compare1} has been prepared.}
\begin{center}
\begin{tabular}{|l|c|c|c|c|c|}
\hline                                              & detector                           & dist.   & power & proton driver            & years  \\
                                                       & vol. (kt)/type                     & (km)   & (MW)  & energy (GeV) & $\nu/\bar\nu$  \\
\hline ESS$\nu$SB-360            & 500/WC                            & 360     & 5         &   2.0/3.0                  &  2/8        \\
           ESS$\nu$SB-540             & 500/WC                            & 560     & 5         &   2.0/3.0                 &   2/8        \\
\hline Hyper-K~\cite{1209.5973, 1109.3262, 0711.2950}     & 560/WC                            & 295     & 0.75    &   30                   &  3/7      \\
\hline  LBNE-10~\cite{lbne-url, 1110.6249, 1307.7335}     &  10/LAr                              & 1290   & 0.72     &    120              &  5/5       \\
            LBNE-PX                           & 34/LAr                               & 1290   & 2.2     &     120              & 5/5     \\
\hline  LBNO-EoI~\cite{lbno_eoi}     & 20/LAr                             & 2300    & 0.7    &     400                &   5/5   \\
\hline  IDS-NF~\cite{1112.2853, 1208.2735}            & 100/MIND                             & 2000      & 4      &     $10^*$                  &   $10^{**}$  \\
\hline  NuMAX~\cite{1301.7727, 1308.0494}      & 10/LAr (magnetized)       & 1300      & 1      &     $5^*$            &   5/5  \\
\hline
\end{tabular}
\end{center}
\vskip -0.1in
{\small $^*$Muon beam energy, relevant for IDS--NF (Low Energy Neutrino Factory) and NuMax.}\\
{\small $^{**}$IDS-NF is supposed to use at the same time muons and anti--muons.}
\label{experiments}
\end{table}

\subsubsection{Precision on the Leptonic Dirac CP--Violating Phase}

\begin{figure}[hbt]
\begin{center}
 \includegraphics[width=0.6\textwidth]{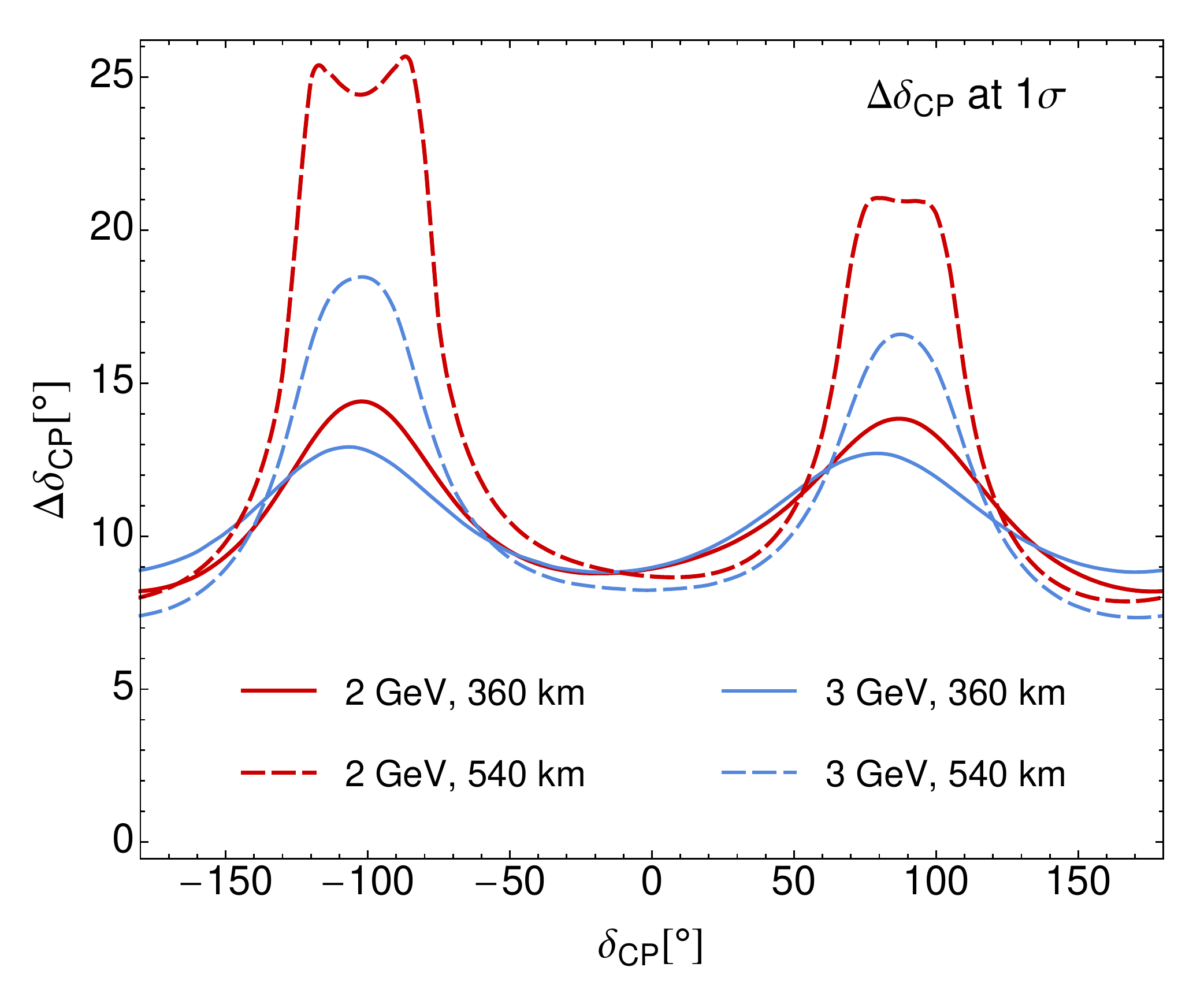}
\end{center}
\caption{The expected 1~$\sigma$ error in $\delta_{CP}$, $\Delta\delta_{CP}$, versus $\delta_{CP}$ for various proton energies and baselines for the ESS Neutrino Super Beam.}
 \label{precision}
\end{figure}

After discovering  leptonic CP violation (i.e $\delta_{CP}\neq 0^\circ,\ 180^\circ$) the next goal will be to measure $\delta_{CP}$ (now the only remaining unmeasured PMNS mixing matrix parameter) as precisely as possible.
The precision that can be reached in this measurement represents an additional discriminating criterion among the experimental options.
Fig.~\ref{precision} presents the precision on $\delta_{CP}$ ($\Delta\delta_{CP}$) versus $\delta_{CP}$.
From this figure one can observe that going to higher proton energies the precision on $\delta_{CP}$ significantly improves in the regions $\pm 45^\circ$ to $\pm 135^\circ$ but has only little variation around $0^\circ$ or $180^\circ$, which are the values that are critical for CP violation discovery.
For 2.0~GeV proton energy, the best baseline for CP violation discovery is for distances from 300~km to 550~km.

\begin{figure}[hbt]
\begin{center}
\includegraphics[width=0.49\textwidth]{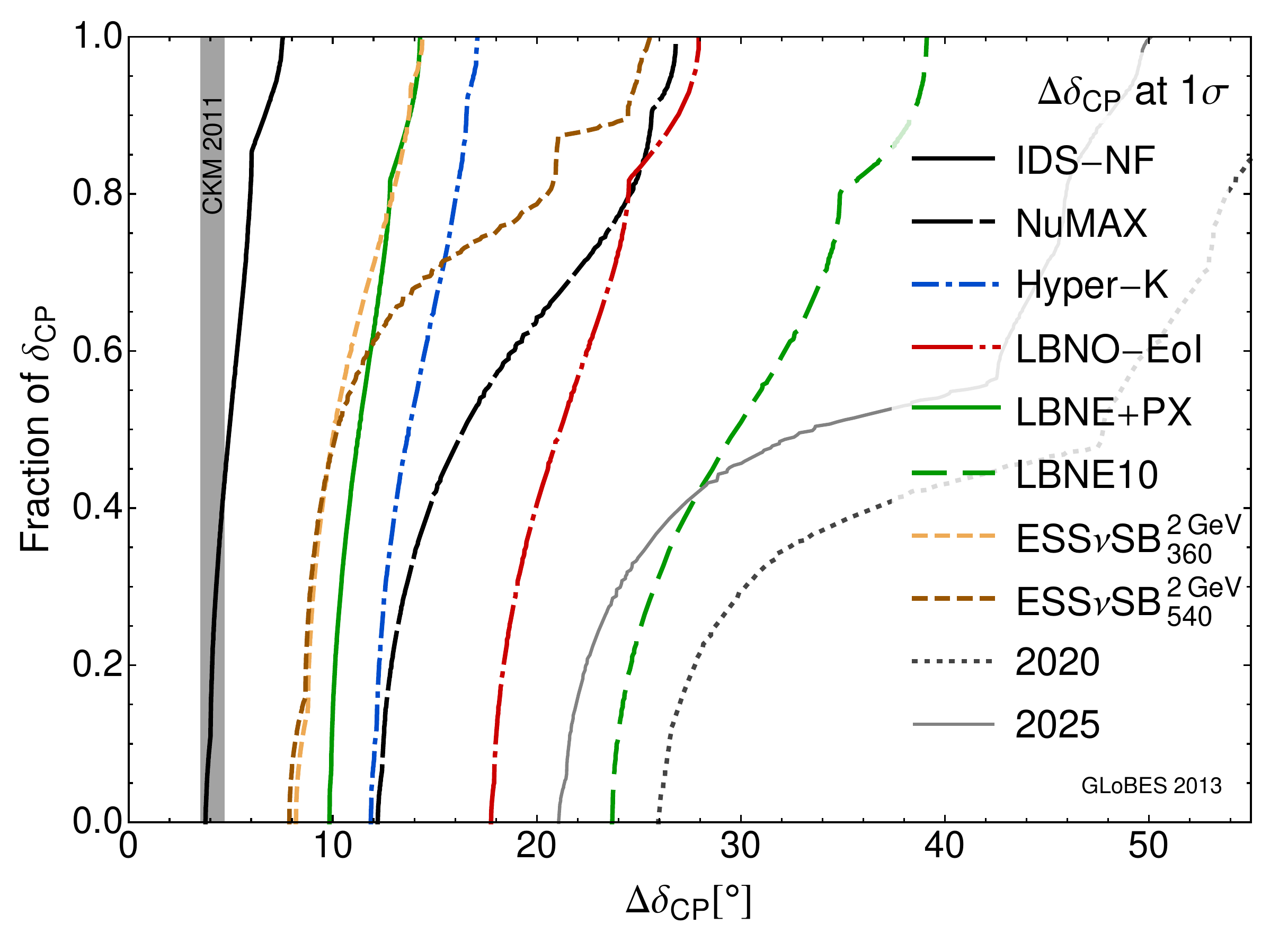}
\includegraphics[width=0.49\textwidth]{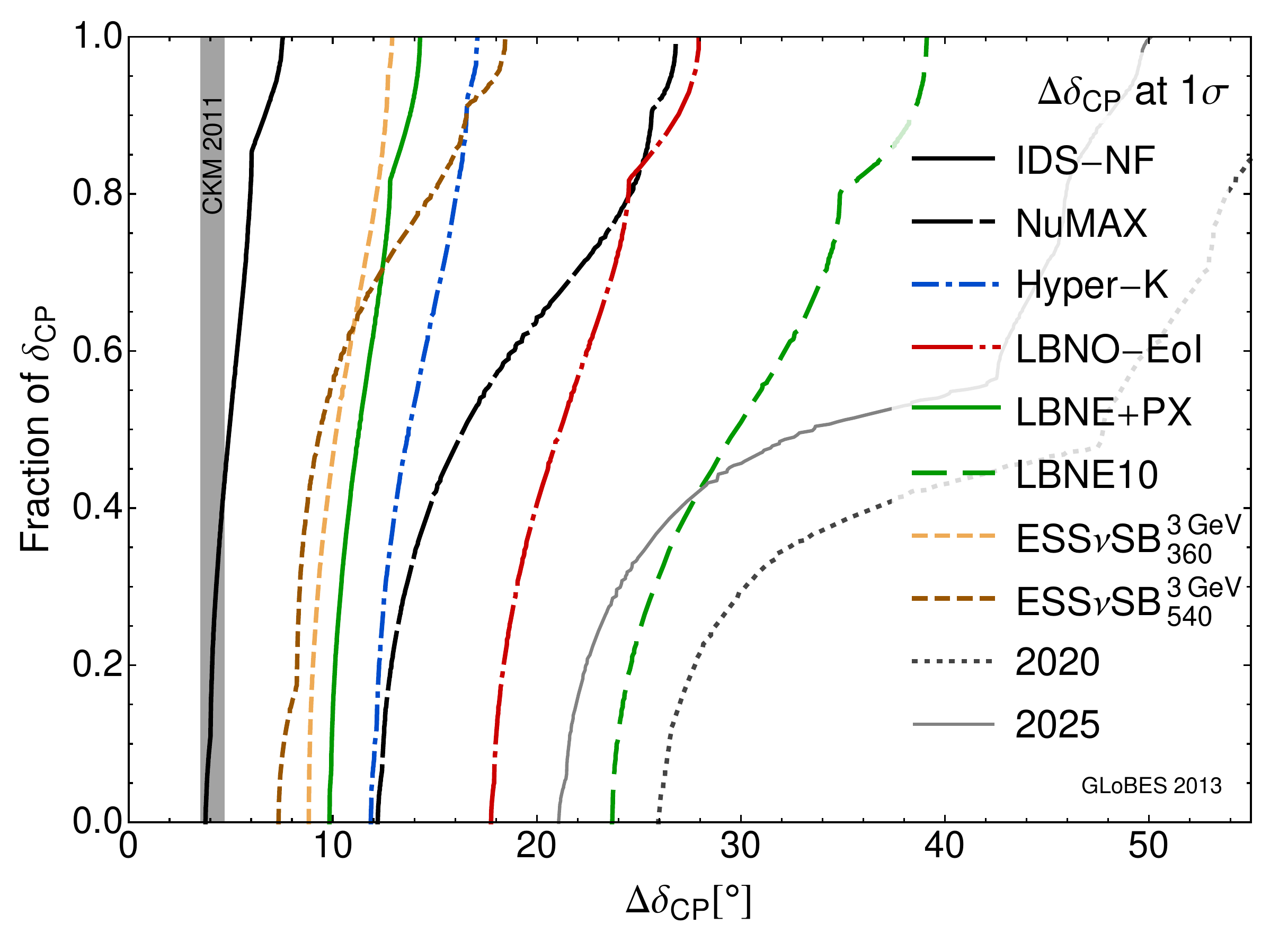}
\end{center}
\caption{Fraction of the full~$\delta_{CP}$ range for which an error $\Delta\delta_{CP}$ (at 1~$\sigma$, 1~d.o.f. ) or better could be achieved.
For ESS$\nu$SB the two baselines of 360~km and 540~km and two proton energies (2.0~GeV on left and 3.0~GeV on right) have been used.}
 \label{compare2}
\end{figure}

Fig.~\ref{compare2}~\cite{snowmassref-url} presents a comparison with other projects of the fraction of the full~$\delta_{CP}$ range covered for which an error $\Delta{\delta_{CP}}$ or better could be achieved.
 Here also it can be seen that only the Neutrino Factory would have a better performance than ESS$\nu$SB.

\subsection{Mass Hierarchy}

\begin{figure}[hbt]
\begin{center}
 \includegraphics[width=0.6\textwidth]{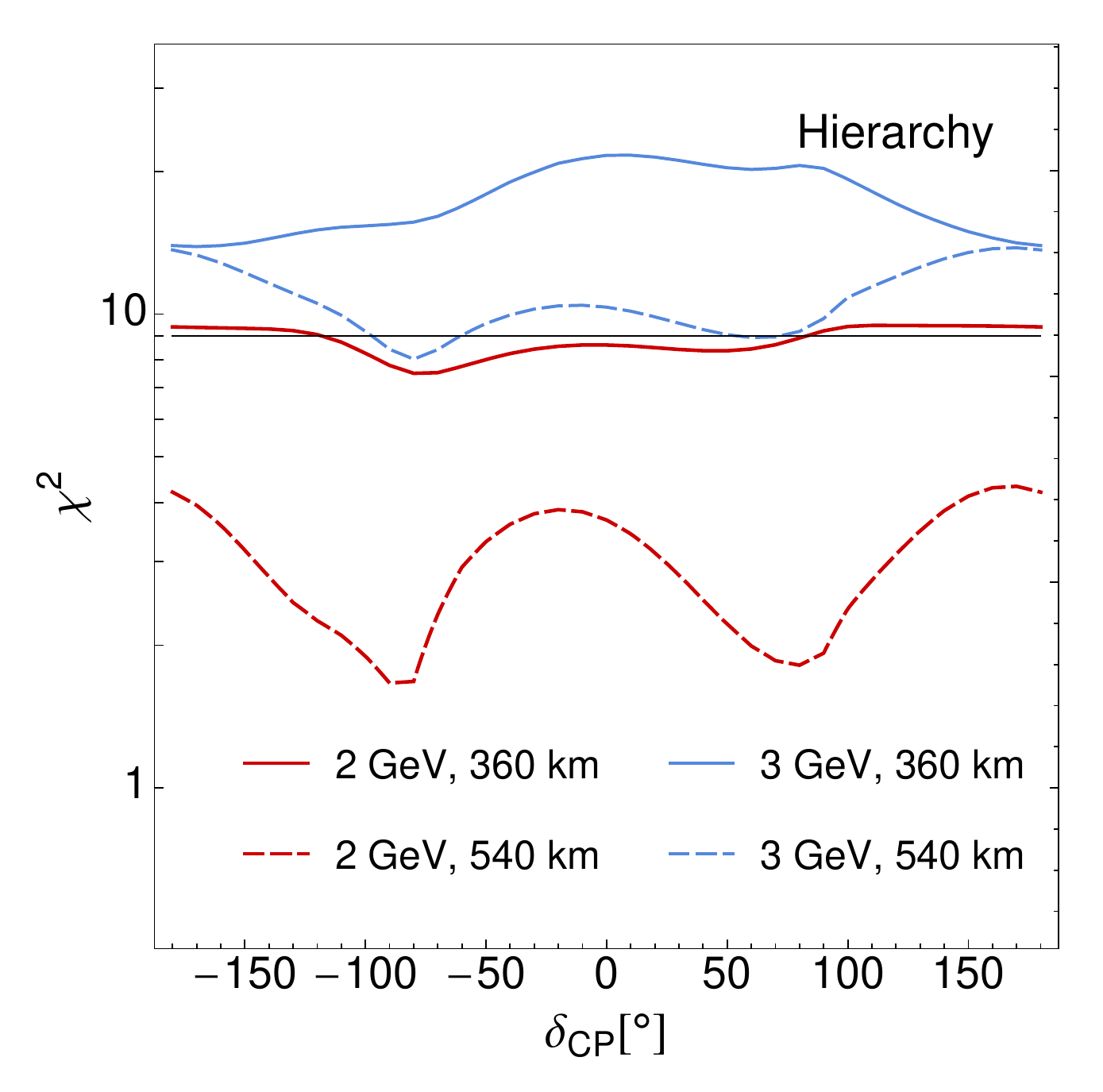}
\end{center}
\caption{The significance in terms of $\sigma$ of the determination of the neutrino mass hierarchy as function of the fraction of the full $\delta_{CP}$ range from -180$^{\circ}$ to 180$^{\circ}$ (atmospheric neutrino data are not included).}
 \label{hierarchy}
\end{figure}

To investigate the significance with which the neutrino mass hierarchy can be determined the analysis was carried out for a series of~$\delta_{CP}$ values between -180$^{\circ}$ and 180$^{\circ}$ assuming a true normal hierarchy and evaluating the $\chi^2$ that would be obtained for the inverted hierarchy.
The results for the opposite choice of true hierarchy are very similar.
The result is shown in Fig.~\ref{hierarchy}.
According to these results the mass hierarchy may be determined at 3~$\sigma$ significance (with the assumption that $n\sigma=\sqrt{\chi^2}$~\cite{1210.3651}) within a large part of the full~$\delta_{CP}$ range for baselines 360~km and 540~km (for 3.0~GeV) but to reach 5~$\sigma$ significance the proton energy has to exceed 3.0~GeV.
However, if combining these Super Beam neutrino measurements with the atmospheric neutrino measurements that can be made concurrently with the same detector \cite{hep-ph/0603172}, it should be possible - owing to the very large volume of the MEMPHYS detector - to obtain results for the mass hierarchy of higher significance than shown in Fig.~\ref{hierarchy}.
As there are reasons to believe that the mass hierarchy ambiguity will be resolved by then (see for example~\cite{1306.1423, 1212.1305, 1203.3388, 1303.6733, reno-url, 1307.7419, 1306.5846, 1306.3988, 1111.4483}), if any compromise has to be done during this optimization phase, CP violation discovery will be privileged.

\section{Available Mines}

\begin{figure}[hbt]
\begin{center}
  \includegraphics[width=0.65\textwidth]{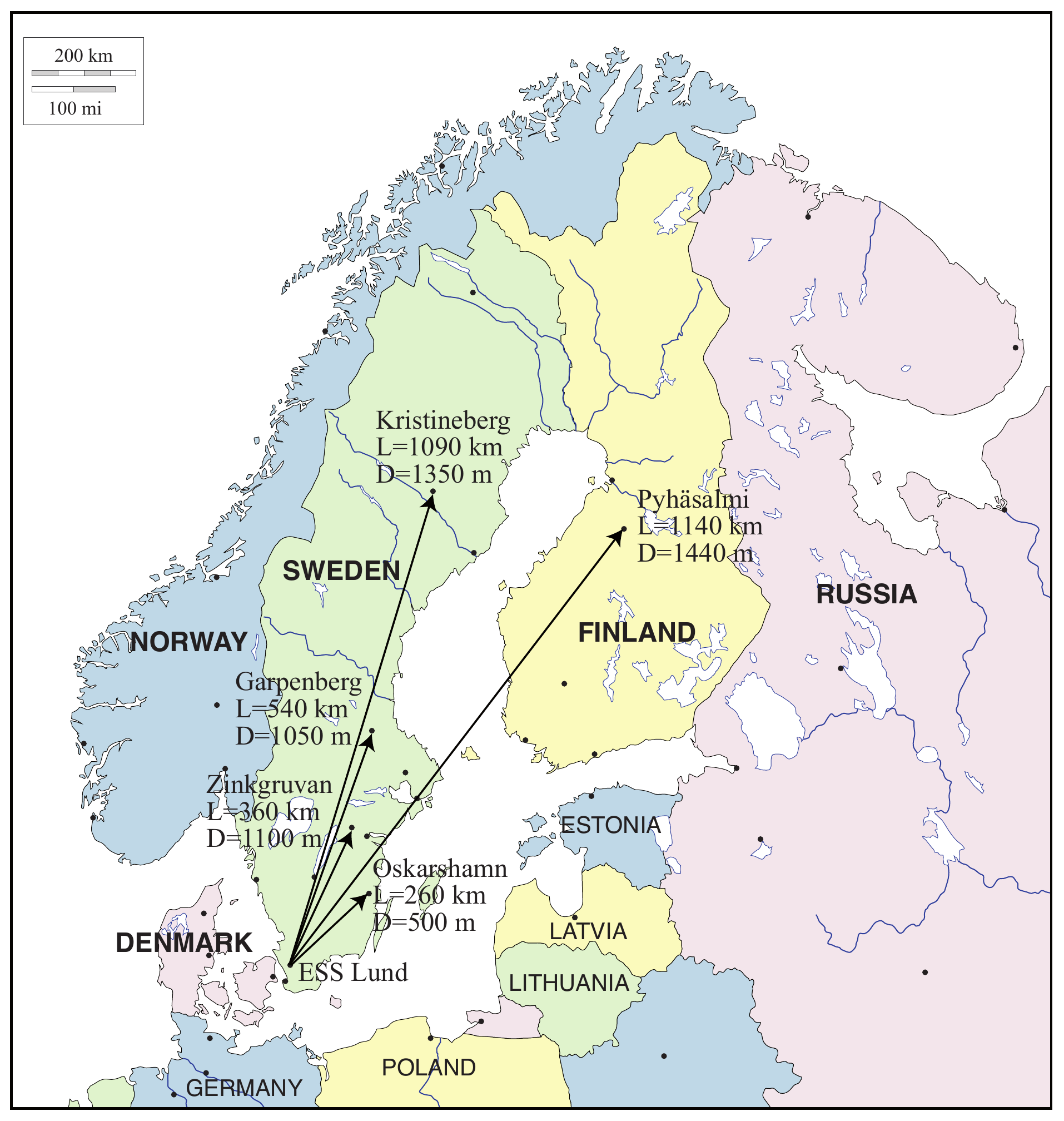}
\end{center}
\caption{ Location of the ESS site and the sites of several deep ($>$ 1000 m) underground mines. The distance ($L$) from ESS--Lund and the depth ($D$) of each mine is indicated below the name of the mine.}
\label{mines}
\end{figure}

In order to protect the large neutrino detector from the cosmic ray background, the far detector needs to be located underground at a level of about 1000~m.
This is important in order to be able to carry out what will be the full research program of a MEMPHYS detector which comprises, in addition to measurements  of Super Beam neutrinos, measurements of proton decay, atmospheric neutrinos, supernovae neutrinos and geoneutrinos.
The total volume of the detector, including the surrounding cosmic--muon veto detector, used for the present calculations is of the order of $6.5\times 10^5$~m$^3$.
The excavation and lining of such a volume in the rock at 1000~m depth level will require the use of a large hoist shaft and a decline down to this level.
Such infrastructure is normally available in mines.
In Section~\ref{detector} the two active mines which, according to our current simulations, are closest to the optimal distance from ESS in Lund, Garpenberg and Zinkgruvan, have already been described .
Fig.~\ref{mines} shows the position in Sweden, the distance (km) from the ESS site in Lund and the depth (m) of these two mines together with several other mines that are (or could be made) deeper than 1000~m.

Among these other mines the closest inactive mine is St\"allberg, which is 1050~m deep and situated 490~km from Lund.
The exploitation of this mine ceased only about 35 years ago but its infrastructure could possibly be reactivated. 
However, it is considered that it is an advantage for the installation and operation of the neutrino detector if the host  mine is an active mine with all its  infrastructures in an operational state.
Another, even nearer underground site is offered by the Oskarhamn nuclear waste depository site, which is situated at a distance of 260~km.
This depository is only 500~m deep, but can be extended down to below 1000~m.
There are more active and inactive mines further north in Sweden. Among the active mines there are Renstr\"om 1240~m deep and Kristineberg 1350~m deep, both at 1090~km from Lund.
The Pyh\"asalmi mine in Finland, which has been studied as detector site for a long baseline beam from the SPS accelerator at CERN, is 1440~m deep and situated 1140~km from Lund.
It is clear, however, that these more northern mines are too far from ESS in Lund to be of primary interest for the ESS neutrino beam project.

\section{Summary and Conclusions}

The currently planned and approved European Spallation Source linac will start delivering protons in 2019.
Providing it with an extra H$^-$ source, an additional 5~MW radiofrequency power source, an accumulator ring, a neutrino target with horn and a decay tunnel, would make possible the production of a neutrino beam of about 300~MeV mean energy derived from~$2.7\times 10^{23}$, 2.0~GeV protons on target per year in concurrent operation with spallation neutron production.
The investment cost for upgrading the ESS linac to produce an extra 5~MW beam is significantly lower than the cost to build a new separate proton driver of the same power.

Such a neutrino beam has the potential to become, during the next decade, the most intense neutrino beam in Europe and maybe also in the world.
The present preliminary study is based on Monte Carlo simulation of the generation and $\nu_\mu \to \nu_e$ oscillation of an initially almost pure $\nu_\mu$ beam produced with the use of the ESS proton linac and of the detection of $\nu_e$ using a very large Water Cherenkov detector of the MEMPHYS type. 
In this study has been determined the range in which the Dirac CP violating phase $\delta_{CP}$ in the lepton sector could be discovered.
The preferred range of distances from the neutrino source to the detector site, within which a comparatively high potential for CP violation discovery is found, is between 300~km and 550~km.
The results indicate that with eight years of data taking with an anti--neutrino beam and two years with a neutrino beam up to 50\% (74\%) of the total CP violation~$\delta_{CP}$ phase range could be covered at 5~$\sigma$ (3~$\sigma$) level at the optimal baseline of around 500~km 
(for these results systematic errors compatible with the {\it snowmass} 2013 process assumptions have been considered).
This coverage, for the EURO$\nu$--like studies where 5\%/10\% systematic errors have been considered for signal/background, go up to 59\% (78\%) at 5~$\sigma$ (3~$\sigma$) confidence level.
With the same baseline, the neutrino mass hierarchy could be determined at more than 3~$\sigma$ level over most of the total~$\delta_{CP}$ phase range without extra optimization and without taking into account atmospheric neutrinos that would improve significantly this performance.

There are several underground mines with a depth of more than 1000~m, which could be used to facilitate the creation of the underground site for the neutrino detector and which are situated within or near the optimal baseline range.
Further optimization of the systematic uncertainties and, e.g., of the target station parameters and also of the detector design, are planned to improve on the sensitivity of the measurements of leptonic CP violation and to progress further with the determination of the optimal baseline.

\section{Acknowledgements}

We are indebted to the European Spallation Source AB for the support we have received.
We wish to thank engineering staff at the Garpenberg and Zinkgruvan mines in Sweden for valuable technical advice and information regarding these respective mines.
This work has been partially supported by the U.S. Department of Energy under award number DE-SC0003915.
P.~Coloma would also like to thank R.~Patterson and R.~Rameika for providing the NOvA simulation details, and M.~Bass, M.~Bishai and E.~Worcester for providing simulation details for LBNE.
E.~Fernandez~Martinez acknowledges financial support by the European Union through the FP7 Marie Curie Actions CIG NeuProbes (PCIG11-GA-2012-321582) and the ITN INVISIBLES (PITN-GA-2011-289442), and the Spanish MINECO through the ``Ram\'on y Cajal'' programme (RYC2011-07710), through the project FPA2009-09017 and through the ``Centro de Excelencia Severo Ochoa'' programme under grant SEV-2012-0249.

\end{document}